\documentclass[journal]{IEEEtran}%
\pdfoutput=1
\usepackage{amsfonts}
\usepackage{amsmath}
\usepackage{amssymb}
\usepackage{graphicx}
\usepackage{verbatim}%
\providecommand{\U}[1]{\protect\rule{.1in}{.1in}}
\input{Qcircuit}

\newtheorem{definition}{Definition}
\newtheorem{example}{Example}

\newtheorem{remark}{Remark}

\begin{document}

\title{Convolutional Entanglement Distillation}
\author{Mark~M.~Wilde, Hari~Krovi,~and~Todd~A.~Brun\thanks{Mark M. Wilde, Hari Krovi,
and Todd A. Brun are with the Communication Sciences Institute of the Ming
Hsieh Department of Electrical Engineering at the University of Southern
California, Los Angeles, California 90089 USA (E-mail: mark.wilde@usc.edu;
krovi@usc.edu; tbrun@usc.edu).}}
\maketitle

\begin{abstract}
We develop a theory of entanglement distillation that exploits a convolutional
coding structure. We provide a method for converting an arbitrary classical
binary or quaternary convolutional code into a convolutional entanglement
distillation protocol. The imported classical convolutional code does not have
to be dual-containing or self-orthogonal. The yield and error-correcting
properties of such a protocol depend respectively on the rate and
error-correcting properties of the imported classical convolutional code. A
convolutional entanglement distillation protocol has several other benefits.
Two parties sharing noisy ebits can distill noiseless ebits \textquotedblleft
online\textquotedblright\ as they acquire more noisy ebits. Distillation yield
is high and decoding complexity is simple for a convolutional entanglement
distillation protocol. Our theory of convolutional entanglement distillation
reduces the problem of finding a good convolutional entanglement distillation
protocol to the well-established problem of finding a good classical
convolutional code.

\end{abstract}

\begin{IEEEkeywords}
quantum convolutional codes, convolutional entanglement distillation, quantum
information theory, entanglement-assisted quantum codes, catalytic codes
\end{IEEEkeywords}

\section{Introduction}

The theory of quantum error correction
\cite{PhysRevA.52.R2493,PhysRevA.54.1098,PhysRevLett.77.793,thesis97gottesman,ieee1998calderbank}%
\ plays a prominent role in the practical realization and engineering of
quantum computing and communication devices. The first quantum
error-correcting codes
\cite{PhysRevA.52.R2493,PhysRevA.54.1098,PhysRevLett.77.793,ieee1998calderbank}%
\ are strikingly similar to classical block codes \cite{book1983code} in their
operation and performance. Quantum error-correcting codes restore a noisy,
decohered quantum state to a pure quantum state. A \textit{stabilizer}
\cite{thesis97gottesman}\ quantum error-correcting code appends ancilla qubits
to qubits that we want to protect. A unitary encoding circuit rotates the
global state into a subspace of a larger Hilbert space. This highly entangled,
encoded state corrects for local noisy errors.
Figure~\ref{fig:stabilizer-code} illustrates the above procedures for encoding
a stabilizer code. A quantum error-correcting code makes quantum computation
and quantum communication practical by providing a way for a sender and
receiver to simulate a noiseless qubit channel given a noisy qubit channel
that has a particular error model.

The stabilizer theory of quantum error correction allows one to import some
classical binary or quaternary codes for use as a quantum code. The only
\textquotedblleft catch\textquotedblright\ when importing is that the
classical code must satisfy the dual-containing or self-orthogonality
constraint. Researchers have found many examples of classical codes satisfying
this constraint \cite{ieee1998calderbank}, but most classical codes do not.

Brun, Devetak, and Hsieh extended the standard stabilizer theory of quantum
error correction by developing the entanglement-assisted stabilizer formalism
\cite{science2006brun,arx2006brun}. They included entanglement as a resource
that a sender and receiver can exploit for a quantum error-correcting code.
They provided a \textquotedblleft direct-coding\textquotedblright%
\ construction in which a sender and receiver can use ancilla qubits and
ebits\footnote{An ebit is a nonlocal bipartite Bell state $\left\vert \Phi
^{+}\right\rangle =\left(  \left\vert 00\right\rangle +\left\vert
11\right\rangle \right)  /\sqrt{2}$.} in a quantum code.\ Gottesman later
showed that their construction is optimal \cite{unpub2007got}---it gives the
minimum number of ebits required for the entanglement-assisted quantum code.
The benefit of including shared entanglement is that one can import an
arbitrary classical binary or quaternary code for use as an
entanglement-assisted quantum code. Another benefit of using shared
entanglement, in addition to being able to import an arbitrary classical
linear code, is that the performance of the original classical code determines
the performance of the resulting quantum code. The entanglement-assisted
stabilizer formalism thus is a significant and powerful extension of the
stabilizer formalism.

The goal of entanglement distillation resembles the goal of quantum error
correction \cite{PhysRevLett.76.722,PhysRevA.54.3824}. An entanglement
distillation protocol extracts noiseless, maximally-entangled ebits from a
larger set of noisy ebits. A sender and receiver can use these noiseless ebits
as a resource for several quantum communication protocols
\cite{PhysRevLett.69.2881,PhysRevLett.70.1895}.

Bennett et al. showed that a strong connection exists between quantum
error-correcting codes and entanglement distillation and demonstrated a method
for converting an arbitrary quantum error-correcting code into a one-way
entanglement distillation protocol \cite{PhysRevA.54.3824}. A one-way
entanglement distillation protocol utilizes one-way classical communication
between sender and receiver to carry out the distillation procedure. Shor and
Preskill\ improved upon Bennett et al.'s method by avoiding the use of ancilla
qubits and gave a simpler method for converting an arbitrary CSS quantum
error-correcting code into an entanglement distillation protocol
\cite{PhysRevLett.85.441}. Nielsen and Chuang showed how to convert a
stabilizer quantum error-correcting code into a stabilizer entanglement
distillation protocol \cite{book2000mikeandike}. Luo and Devetak then
incorporated shared entanglement to demonstrate how to convert an
entanglement-assisted stabilizer code into an entanglement-assisted
entanglement distillation protocol \cite{luo:010303}. All of the above
constructions exploit the relationship between quantum error correction and
entanglement distillation---we further exploit the connection in this paper by
forming a \textit{convolutional} entanglement distillation protocol.

Several authors have recently contributed toward a theory of quantum
convolutional codes
\cite{PhysRevLett.91.177902,arxiv2004olliv,isit2005forney,ieee2007forney}.
Quantum convolutional codes are useful in a communication context where a
sender has a large stream of qubits to send to a receiver. Quantum
convolutional codes are similar to classical convolutional codes in their
operation and performance \cite{arxiv2004olliv,ieee2007forney}. Classical
convolutional codes have some advantages over classical block codes such as
superior code rates and lower decoding complexity \cite{book1999conv}. Their
quantum counterparts enjoy these same advantages over quantum block codes
\cite{ieee2007forney}.

The development of quantum convolutional codes has been brief but successful.
Chau was the first to construct some quantum convolutional codes
\cite{PhysRevA.58.905,PhysRevA.60.1966}, though some authors
\cite{ieee2007forney}\ argue that his construction is not a true quantum
convolutional code. Several authors have established a working theory of
quantum convolutional coding based on the stabilizer formalism and classical
self-orthogonal codes over the finite field $\mathbb{F}_{4}$
\cite{PhysRevLett.91.177902,arxiv2004olliv,isit2005forney,ieee2007forney}.
Others have also provided a practical way for realizing \textquotedblleft
online\textquotedblright\ encoding and decoding circuits for quantum
convolutional codes
\cite{PhysRevLett.91.177902,arxiv2004olliv,ieee2006grassl,isit2006grassl}.
These successes have led to a theory of quantum convolutional coding which is
useful but not complete. We add to the usefulness of the quantum convolutional
theory by considering entanglement distillation and shared entanglement.

In this paper, our main contribution is a theory of convolutional entanglement
distillation. Our theory allows us to import the entirety of classical
convolutional coding theory for use in entanglement distillation. The task of
finding a good convolutional entanglement distillation protocol now becomes
the well-established task of finding a good classical convolutional code.

We begin in Section~\ref{sec:conv-ent-dist} by showing how to construct
a\textit{ }convolutional entanglement distillation protocol from an arbitrary
quantum convolutional code. We translate earlier protocols
\cite{PhysRevLett.85.441,book2000mikeandike}\ for entanglement distillation of
a block of noisy ebits to the convolutional setting. Our convolutional
entanglement distillation protocol possesses several benefits---it has a
higher distillation yield and lower decoding complexity than a block
entanglement distillation protocol. A convolutional entanglement distillation
protocol has the additional benefit of distilling entanglement
\textquotedblleft online.\textquotedblright\ This online property is useful
because the sender and receiver can distill entanglement \textquotedblleft on
the fly\textquotedblright\ as they obtain more noisy ebits. This translation
from a quantum convolutional code to an entanglement distillation protocol is
useful because it paves the way for our major contribution.

Our major advance is a method for constructing a convolutional entanglement
distillation protocol when the sender and receiver initially share some
noiseless ebits. All prior quantum convolutional work requires the code to
satisfy the restrictive self-orthogonality constraint, and authors performed
specialized searches for classical convolutional codes that meet this
constraint
\cite{PhysRevLett.91.177902,arxiv2004olliv,isit2005forney,ieee2007forney}. We
lift this constraint by allowing shared noiseless entanglement. The benefit of
convolutional entanglement distillation with entanglement assistance is that
we can import an \textit{arbitrary} classical binary or quaternary
convolutional code for use in a convolutional entanglement distillation
protocol. The error-correcting properties for the convolutional entanglement
distillation protocol follow directly from the properties of the imported
classical code. Thus we can apply the decades of research on classical
convolutional coding theory with many of the benefits of the convolutional
structure carrying over to the quantum domain.

We organize our work as follows. In Section \ref{sec:stabilizer}, we review
the stabilizer theory for quantum error correction and entanglement
distillation. The presentation of the mathematics is similar in style to Refs.
\cite{ieee2007forney,arx2006brun}. The stabilizer review includes a review of
the standard stabilizer theory (Section \ref{sec:standard-stabilizer}), the
entanglement-assisted stabilizer theory\ (Section
\ref{sec:entang-assisted-stabilizer}), convolutional stabilizer codes (Section
\ref{sec:conv-stabilizer}), stabilizer entanglement distillation (Section
\ref{sec:stabilizer-ent-distill}), and entanglement-assisted entanglement
distillation (Section \ref{sec:stabilizer-ent-assist-ent-distill}). We provide
a small contribution in the Appendix---a simple algorithm to determine an
encoding circuit and the optimal number of ebits required for an
entanglement-assisted block code. The original work \cite{arx2006brun}\ gave
two theorems relevant to the encoding circuit, but the algorithm we present
here is simpler. In Section \ref{sec:conv-ent-dist}, we show how to convert an
arbitrary quantum convolutional code into a convolutional entanglement
distillation protocol. In Section \ref{sec:conv-ent-ent-assist}, we provide
several methods and examples for constructing convolutional entanglement
distillation protocols where two parties possess a few initial noiseless
ebits. These initial noiseless ebits act as a catalyst for the convolutional
distillation protocol. The constructions in Section
\ref{sec:conv-ent-ent-assist}\ make it possible to import an arbitrary
classical binary or quaternary convolutional code for use in convolutional
entanglement distillation.

\section{Review of the Stabilizer Formalism}

\label{sec:stabilizer}

\subsection{Standard Stabilizer Formalism for Quantum Block Codes}

\label{sec:standard-stabilizer}The stabilizer formalism exploits elements of
the Pauli group $\Pi$ in formulating quantum error-correcting codes. The set
$\Pi=\left\{  I,X,Y,Z\right\}  $ consists of the Pauli operators:%
\[
I\equiv%
\begin{bmatrix}
1 & 0\\
0 & 1
\end{bmatrix}
,\ X\equiv%
\begin{bmatrix}
0 & 1\\
1 & 0
\end{bmatrix}
,\ Y\equiv%
\begin{bmatrix}
0 & -i\\
i & 0
\end{bmatrix}
,\ Z\equiv%
\begin{bmatrix}
1 & 0\\
0 & -1
\end{bmatrix}
.
\]
The above operators act on a single qubit---a state in a two-dimensional
Hilbert space. Operators in $\Pi$ have eigenvalues $\pm1$ and either commute
or anti-commute. The set $\Pi^{n}$ consists of $n$-fold tensor products of
Pauli operators:%
\begin{equation}
\Pi^{n}=\left\{
\begin{array}
[c]{c}%
e^{i\phi}A_{1}\otimes\cdots\otimes A_{n}:\forall j\in\left\{  1,\ldots
,n\right\}  ,\\
A_{j}\in\Pi,\ \ \phi\in\left\{  0,\pi/2,\pi,3\pi/2\right\}
\end{array}
\right\}  .
\end{equation}
Elements of $\Pi^{n}$\ act on a quantum register of $n$ qubits. We
occasionally omit tensor product symbols in what follows so that $A_{1}\cdots
A_{n}\equiv A_{1}\otimes\cdots\otimes A_{n}$. The $n$-fold Pauli group
$\Pi^{n}$\ plays an important role for both the encoding circuit and the
error-correction procedure of a quantum stabilizer code over $n$ qubits.

\begin{figure}
[ptb]
\begin{center}
\includegraphics[
natheight=3.386600in,
natwidth=8.973300in,
height=1.2038in,
width=3.1747in
]%
{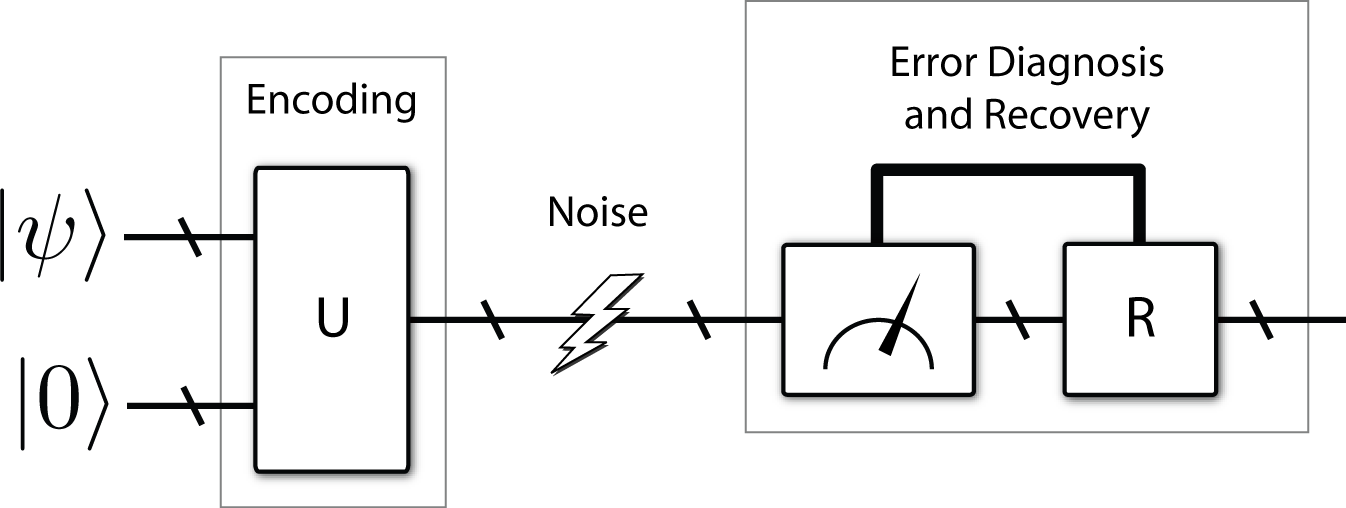}%
\caption{The operation of a stabilizer code. Thin lines denote quantum
information and thick lines denote classical information. Slanted bars denote
multiple qubits. A sender encodes a multi-qubit state $\left\vert
\psi\right\rangle $ with the help of some ancilla qubits $\left\vert
0\right\rangle $. She sends the encoded state over a noisy quantum channel.
The receiver performs multi-qubit measurements to extract information about
the errors. He finally performs a recovery operation $R$\ to reverse the
channel error.}%
\label{fig:stabilizer-code}%
\end{center}
\end{figure}

Let us define an $\left[  n,k\right]  $\ stabilizer quantum error-correcting
code to encode $k$ logical qubits into $n$ physical qubits. The rate of such a
code is $k/n$. Its stabilizer $\mathcal{S}$\ is an abelian subgroup of the
$n$-fold Pauli group $\Pi^{n}$: $\mathcal{S}\subset\Pi^{n}$. $\mathcal{S}$
does not contain the operator $-I^{\otimes n}$. The simultaneous
$+1$-eigenspace of the operators constitutes the \textit{codespace}. The
codespace has dimension $2^{k}$ so that we can encode $k$\ qubits into it. The
stabilizer $\mathcal{S}$ has a minimal representation in terms of $n-k$
independent generators $\left\{  g_{1},\ldots,g_{n-k}\ |\ \forall i\in\left\{
1,\ldots,n-k\right\}  ,\ g_{i}\in\mathcal{S}\right\}  $. The generators are
independent in the sense that none of them is a product of any other two (up
to a global phase). The operators $g_{1},\ldots,g_{n-k}$ function in the same
way as a parity check matrix does for a classical linear block code.
Figure~\ref{fig:stabilizer-code} illustrates the operation of a stabilizer
code.%

One of the fundamental notions in quantum error correction theory is that it
suffices to correct a discrete error set with support in the Pauli group
$\Pi^{n}$ \cite{PhysRevA.52.R2493}. Suppose that the errors affecting an
encoded quantum state are a subset $\mathcal{E}$ of the Pauli group $\Pi^{n}$:
$\mathcal{E}\subset\Pi^{n}$. An error $E\in\mathcal{E}$ that affects an
encoded quantum state either commutes or anticommutes with any particular
element $g\ $in $\mathcal{S}$. The error $E$\ is correctable if it
anticommutes with an element $g\ $in $\mathcal{S}$. An anticommuting error
$E$\ is detectable by measuring each element $g$\ in $\mathcal{S}$ and
computing a syndrome $\mathbf{r}$ identifying $E$. The syndrome is a binary
vector $\mathbf{r}$\ with length $n-k$ whose elements identify whether the
error $E$ commutes or anticommutes with each $g\in\mathcal{S}$. An error
$E$\ that commutes with every element $g$ in $\mathcal{S}$\ is correctable if
and only if it is in $\mathcal{S}$. It corrupts the encoded state if it
commutes with every element of $\mathcal{S}$ but does not lie in $\mathcal{S}%
$. So we compactly summarize the stabilizer error-correcting conditions: a
stabilizer code can correct any errors $E_{1},E_{2}$ in $\mathcal{E}$ if
$E_{1}^{\dag}E_{2}\notin\mathcal{Z}\left(  \mathcal{S}\right)  $ or
$E_{1}^{\dag}E_{2}\in\mathcal{S}$ where $\mathcal{Z}\left(  \mathcal{S}%
\right)  $ is the centralizer of $\mathcal{S}$.

\subsection{Entanglement-Assisted Stabilizer Formalism}

\label{sec:entang-assisted-stabilizer}
\begin{figure}
[ptb]
\begin{center}
\includegraphics[
natheight=4.266100in,
natwidth=10.639800in,
height=1.3785in,
width=3.4238in
]%
{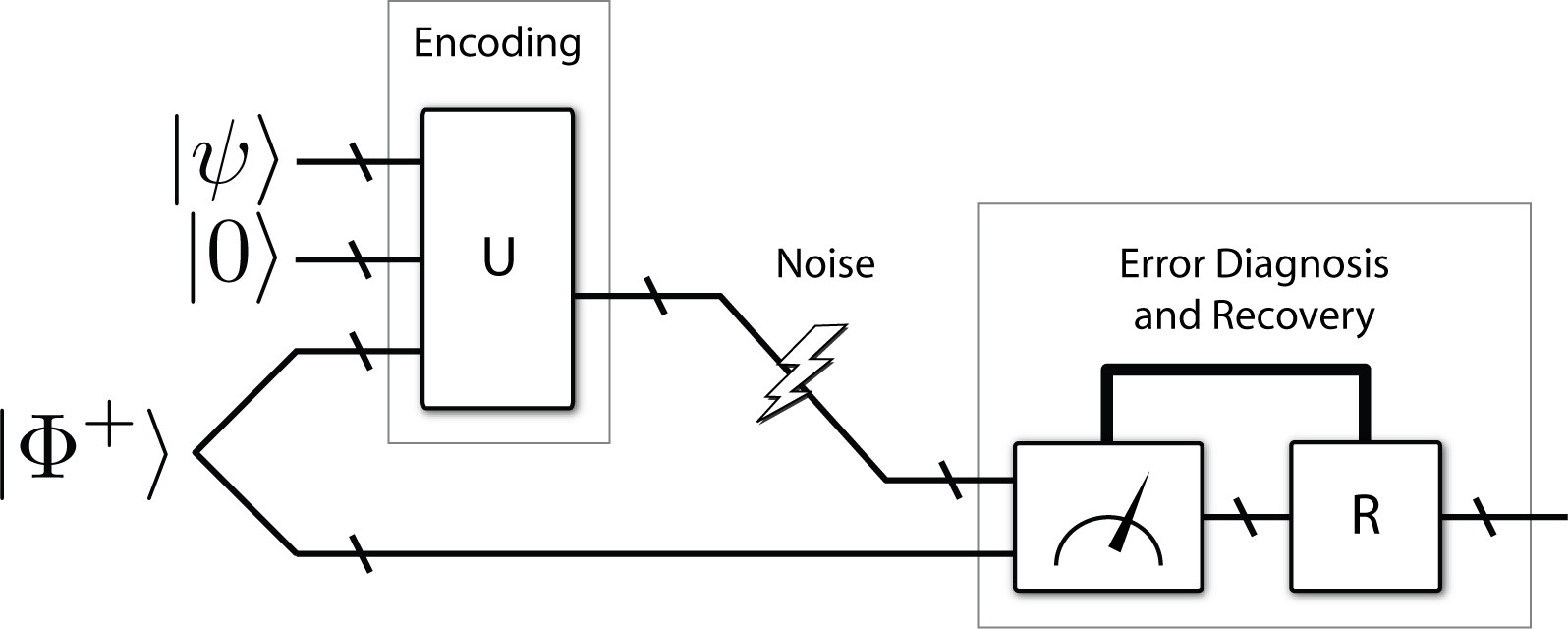}%
\caption{The operation of an entanglement-assisted quantum error-correcting
code. The sender encodes quantum information in state $\left\vert
\psi\right\rangle $\ with the help of local ancilla qubits $\left\vert
0\right\rangle $\ and her half of a set of shared ebits $\left\vert \Phi
^{+}\right\rangle $. She then sends her qubits over a noisy quantum channel.
The channel does not corrupt the receiver's half of the set of shared ebits.
The receiver performs multi-qubit measurements on all of the qubits to
diagnose the channel error. He performs a recovery unitary $R$ to reverse the
channel error.}%
\label{fig:entanglement-assisted-code}%
\end{center}
\end{figure}

The entanglement-assisted stabilizer
formalism extends the standard stabilizer formalism by including shared
entanglement \cite{science2006brun,arx2006brun}.
Figure~\ref{fig:entanglement-assisted-code}\ demonstrates the operation of a
generic entanglement-assisted stabilizer code.

The advantage of entanglement-assisted stabilizer codes is that the sender can
exploit the error-correcting properties of an arbitrary set of Pauli
operators. The sender's Pauli operators do not necessarily have to form an
abelian subgroup of $\Pi^{n}$. The sender can make clever use of her shared
ebits so that the global stabilizer is abelian and thus forms a valid quantum
error-correcting code.%

We review the construction of an entanglement-assisted code. Suppose that
there is a nonabelian subgroup $\mathcal{S}\subset\Pi^{n}$ of size $n-k=2c+s$.
Application of the fundamental theorem of symplectic geometry\footnote{We
loosely refer to this theorem as the fundamental theorem of symplectic
geometry because of its importance in symplectic geometry and in quantum
coding theory.} \cite{book2001symp}\ (Lemma~1 in \cite{science2006brun}%
)\ states that there exists a minimal set\ of independent generators $\left\{
\bar{Z}_{1},\ldots,\bar{Z}_{s+c},\bar{X}_{s+1},\ldots,\bar{X}_{s+c}\right\}  $
for $\mathcal{S}$ with the following commutation relations:%
\begin{align}
\left[  \bar{Z}_{i},\bar{Z}_{j}\right]   &  =0\ \ \ \ \ \forall
i,j,\nonumber\\
\left[  \bar{X}_{i},\bar{X}_{j}\right]   &  =0\ \ \ \ \ \forall
i,j,\nonumber\\
\left[  \bar{X}_{i},\bar{Z}_{j}\right]   &  =0\ \ \ \ \ \forall i\neq
j,\nonumber\\
\left\{  \bar{X}_{i},\bar{Z}_{i}\right\}   &  =0\ \ \ \ \ \forall i.
\end{align}
The decomposition of $\mathcal{S}$ into the above minimal generating set
determines that the code requires $s$ ancilla qubits and $c$ ebits. The code
requires an ebit for every anticommuting pair in the minimal generating set.
The simple reason for this requirement is that an ebit is a simultaneous
$+1$-eigenstate of the operators $\left\{  XX,ZZ\right\}  $. The second qubit
in the ebit transforms the anticommuting pair $\left\{  X,Z\right\}  $ into a
commuting pair $\left\{  XX,ZZ\right\}  $. The above decomposition also
minimizes the number of ebits required for the code \cite{unpub2007got}---it
is an optimal decomposition.

We can partition the nonabelian group $\mathcal{S}$ into two subgroups: the
isotropic subgroup $\mathcal{S}_{I}$\ and the entanglement subgroup
$\mathcal{S}_{E}$. The isotropic subgroup $\mathcal{S}_{I}$ is a commuting
subgroup of $\mathcal{S}$ and thus corresponds to ancilla
qubits:\ $\mathcal{S}_{I}=\left\{  \bar{Z}_{1},\ldots,\bar{Z}_{s}\right\}  $.
The elements of the entanglement subgroup $\mathcal{S}_{E}$ come in
anticommuting pairs and thus correspond to ebits: $\mathcal{S}_{E}=\left\{
\bar{Z}_{s+1},\ldots,\bar{Z}_{s+c},\bar{X}_{s+1},\ldots,\bar{X}_{s+c}\right\}
$.

The two subgroups $\mathcal{S}_{I}$ and $\mathcal{S}_{E}$\ play a role in the
error-correcting conditions for the entanglement-assisted stabilizer
formalism. An entanglement-assisted code corrects errors in a set
$\mathcal{E}\subset\Pi^{n}$ if for all $E_{1},E_{2}\in\mathcal{E}$,
$E_{1}^{\dag}E_{2}\in\mathcal{S}_{I}\cup\left(  \Pi^{n}-\mathcal{Z}\left(
\left\langle \mathcal{S}_{I},\mathcal{S}_{E}\right\rangle \right)  \right)  $.

The operation of an entanglement-assisted code is as follows. The sender
performs an encoding unitary on her unprotected qubits, ancilla qubits, and
her half of the ebits. The unencoded state is a simultaneous +1-eigenstate of
the following operators:%
\begin{equation}
\left\{
\begin{array}
[c]{c}%
Z_{1},\ldots,Z_{s},\\
Z_{s+1}|Z_{1},\ldots,Z_{s+c}|Z_{c},\\
X_{s+1}|X_{1},\ldots,X_{s+c}|X_{c}%
\end{array}
\right\}  .
\end{equation}
The operators to the right of the vertical bars indicate the receiver's half
of the shared ebits. The encoding unitary transforms the unencoded operators
to the following encoded operators:%
\begin{equation}
\left\{
\begin{array}
[c]{c}%
\bar{Z}_{1},\ldots,\bar{Z}_{s},\\
\bar{Z}_{s+1}|Z_{1},\ldots,\bar{Z}_{s+c}|Z_{c},\\
\bar{X}_{s+1}|X_{1},\ldots,\bar{X}_{s+c}|X_{c}%
\end{array}
\right\}  .
\end{equation}
The sender transmits all of her qubits over the noisy quantum channel. The
receiver then possesses the transmitted qubits and his half of the ebits. He
measures the above encoded operators to diagnose the error. The last step is
to correct for the error.

We give an example of an entanglement-assisted stabilizer code in the
Appendix. This example highlights the main features of the theory given above.

The Appendix also gives an original algorithm that determines the encoding
circuit for the sender to perform. The algorithm determines the number of
ancilla qubits and the number of ebits that the code requires.

The defined rate of an entanglement-assisted quantum error-correcting code is
$\left(  k-c\right)  /n$ \cite{science2006brun,arx2006brun}. The authors
defined the rate in this way to compare entanglement-assisted codes with
standard quantum error-correcting codes.

We mention that the rate pair $\left(  k/n,c/n\right)  $ more properly
characterizes the rate of an entanglement-assisted code because an
entanglement-assisted code is a \textquotedblleft father\textquotedblright%
\ code in the sense of Ref. \cite{arx2005dev}. The first number in the pair
gives the rate of noiseless qubits generated per channel use and the second
number gives the rate of ebits consumed per channel use. The rate pair falls
in the two-dimensional capacity region for the \textquotedblleft
father\textquotedblright\ protocol. The goal of an entanglement-assisted
coding strategy is for the rate pair to approach the boundary of the capacity
region as the block length becomes large.

\subsection{Convolutional Stabilizer Codes}

The block codes reviewed above are useful in quantum computing and in quantum
communications. The encoding circuit for a large block code typically has a
high complexity although those for modern codes do have lower complexity.

Quantum convolutional coding theory
\cite{PhysRevLett.91.177902,arxiv2004olliv,isit2005forney,ieee2007forney}
offers a different paradigm for coding quantum information. The convolutional
structure is useful for a quantum communication scenario where a sender
possesses a stream of qubits to send to a receiver. The encoding circuit for a
quantum convolutional code has a much lower complexity than an encoding
circuit needed for a large block code. It also has a repetitive pattern so
that the same physical devices or the same routines can manipulate the stream
of quantum information.

\label{sec:conv-stabilizer}Quantum convolutional stabilizer codes borrow
heavily from the structure of their classical counterparts
\cite{PhysRevLett.91.177902,arxiv2004olliv,isit2005forney,ieee2007forney}.
Quantum convolutional codes are similar because some of the qubits feed back
into a repeated encoding unitary and give the code a memory structure like
that of a classical convolutional code. The quantum codes feature online
encoding and decoding of qubits. This feature gives quantum convolutional
codes both their low encoding and decoding complexity and their ability to
correct a larger set of errors than a block code with similar parameters.

We first review some preliminary mathematics and follow with the definition of
a quantum convolutional stabilizer code \cite{arxiv2004olliv,ieee2007forney}.
We end this section with a brief discussion of encoding circuits for quantum
convolutional codes.

A quantum convolutional stabilizer code acts on a Hilbert space $\mathcal{H}$
that\ is\ a countably infinite tensor product of two-dimensional qubit Hilbert
spaces $\left\{  \mathcal{H}_{i}\right\}  _{i\in\mathbb{Z}^{+}}$ where%
\begin{equation}
\mathcal{H}=%
{\displaystyle\bigotimes\limits_{i=0}^{\infty}}
\ \mathcal{H}_{i}.
\end{equation}
and $\mathbb{Z}^{+}\equiv\left\{  0,1,\ldots\right\}  $. A sequence
$\mathbf{A}$ of Pauli matrices $\left\{  A_{i}\right\}  _{i\in\mathbb{Z}^{+}}%
$, where%
\begin{equation}
\mathbf{A}=%
{\displaystyle\bigotimes\limits_{i=0}^{\infty}}
\ A_{i},
\end{equation}
can act on states in $\mathcal{H}$. Let $\Pi^{\mathbb{Z}^{+}}$ denote the set
of all Pauli sequences. The support supp$\left(  \mathbf{A}\right)  $\ of a
Pauli sequence $\mathbf{A}$ is the set of indices of the entries in
$\mathbf{A}$ that are not equal to the identity. The weight of a sequence
$\mathbf{A}$ is the size $\left\vert \text{supp}\left(  \mathbf{A}\right)
\right\vert $\ of its support. The delay del$\left(  \mathbf{A}\right)  $ of a
sequence $\mathbf{A}$ is the smallest index for an entry not equal to the
identity. The degree deg$\left(  \mathbf{A}\right)  $ of a sequence
$\mathbf{A}$ is the largest index for an entry not equal to the identity.
E.g., the following Pauli sequence%
\begin{equation}%
\begin{array}
[c]{cccccccc}%
I & X & I & Y & Z & I & I & \cdots
\end{array}
,
\end{equation}
has support $\left\{  1,3,4\right\}  $, weight three, delay one, and degree
four. A sequence has finite support if its weight is finite. Let
$F(\Pi^{\mathbb{Z}^{+}})$ denote the set of Pauli sequences with finite
support. The following definition for a quantum convolutional code utilizes
the set $F(\Pi^{\mathbb{Z}^{+}})$ in its description.

\begin{definition}
\label{def:conv-code}A rate $k/n$-convolutional stabilizer code with $0\leq
k\leq n$ is a commuting set $\mathcal{G}$\ of all $n$-qubit shifts of a basic
generator set $\mathcal{G}_{0}$. The basic generator set $\mathcal{G}_{0}$ has
$n-k$ Pauli sequences of finite support:%
\begin{equation}
\mathcal{G}_{0}=\left\{  \mathbf{G}_{i}\in F(\Pi^{\mathbb{Z}^{+}}):1\leq i\leq
n-k\right\}  .
\end{equation}
The constraint length $\nu$ of the code is the maximum degree of the
generators in $\mathcal{G}_{0}$. A frame of the code consists of $n$ qubits.
\end{definition}

\begin{remark}
The above definition requires that all elements of $\mathcal{G}$ commute. In
Section~\ref{sec:conv-ent-ent-assist}, we lift the restrictive commutative
condition when we construct a convolutional entanglement distillation protocol
with entanglement assistance.
\end{remark}

A quantum convolutional code admits an equivalent definition in terms of the
delay transform or $D$-transform. The $D$-transform captures shifts of the
basic generator set $\mathcal{G}_{0}$. Let us define the $n$-qubit delay
operator $D$ acting on any Pauli sequence $\mathbf{A}\in\Pi^{\mathbb{Z}^{+}}%
$\ as follows:%
\begin{equation}
D\left(  \mathbf{A}\right)  =I^{\otimes n}\otimes\mathbf{A.}
\label{eq:delay-transform}%
\end{equation}
We can write $j$ repeated applications of $D$ as a power of $D$:%
\begin{equation}
D^{j}\left(  \mathbf{A}\right)  =I^{\otimes jn}\otimes\mathbf{A.}%
\end{equation}
Let $D^{j}\left(  \mathcal{G}_{0}\right)  $ be the set of shifts of elements
of $\mathcal{G}_{0}$ by $j$. Then the full stabilizer $\mathcal{G}$ for the
convolutional stabilizer code is%
\begin{equation}
\mathcal{G}=%
{\textstyle\bigcup\limits_{j\in\mathbb{Z}^{+}}}
D^{j}\left(  \mathcal{G}_{0}\right)  .
\end{equation}
%

\begin{figure*}
[ptb]
\begin{center}
\includegraphics[
natheight=4.260100in,
natwidth=18.372900in,
height=1.5662in,
width=6.6945in
]
{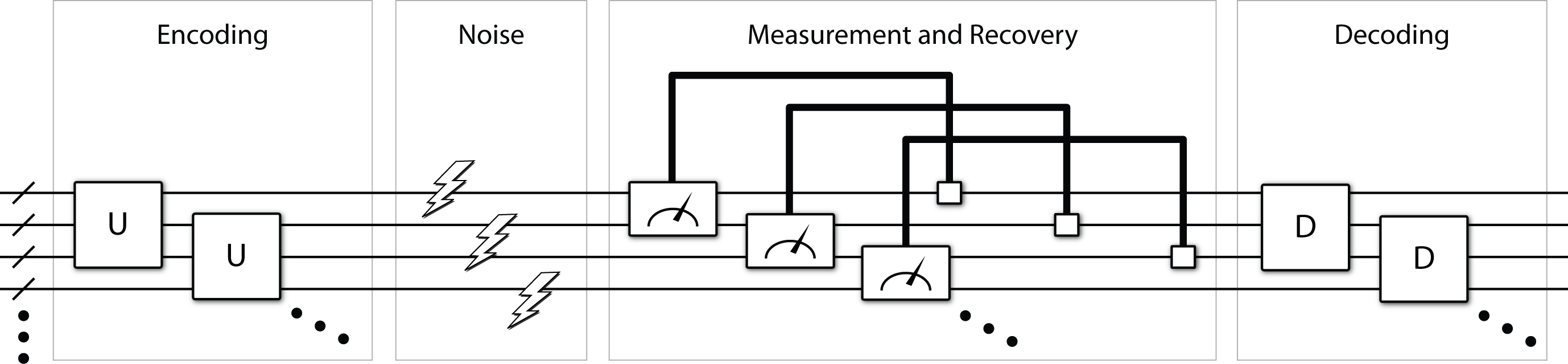}
\caption{An example of a quantum convolutional code.
The sender applies the same unitary successively to a stream of information qubits and ancilla qubits.
The convolutional structure implies that the unitary overlaps some of the same qubits.
The sender transmits her qubits as soon as the unitary finishes processing them.
The noisy quantum channel corrupts the transmitted qubits.
The receiver performs overlapping multi-qubit measurements to diagnose
channel errors and corrects for them. The receiver performs an online decoding circuit to recover
the sender's original stream of information qubits.}
\label{fig:qcc}
\end{center}
\end{figure*}%
Figure~\ref{fig:qcc}\ outlines the operation of a convolutional stabilizer
code. The protocol begins with the sender encoding a stream of qubits with an
online encoding circuit such as that given in \cite{isit2006grassl}. The
encoding circuit is \textquotedblleft online\textquotedblright\ if it acts on
a few blocks of qubits at a time. The sender transmits a set of qubits as soon
as the first unitary finishes processing them. The receiver measures all the
generators in $\mathcal{G}$ and corrects for errors as he receives the online
encoded qubits. He finally decodes the encoded qubits with a decoding circuit.
The qubits decoded from this convolutional procedure should be error free and
ready for quantum computation at the receiving end.

A \textit{finite-depth }circuit maps a Pauli sequence with finite weight to
one with finite weight \cite{arxiv2004olliv}. It does not map a Pauli sequence
with finite weight to one with infinite weight. This property is important
because we do not want the decoding circuit to propagate uncorrected errors
into the information qubit stream \cite{book1999conv}. A finite-depth decoding
circuit corresponding to the stabilizer $\mathcal{G}$ exists by the algorithm
given in \cite{isit2006grassl}.

\begin{example}
\label{sec:qcc-example}Forney et al. provided an example of a rate-1/3 quantum
convolutional code by importing a particular classical quaternary
convolutional code \cite{isit2005forney,ieee2007forney}. Grassl and
R\"{o}tteler determined a noncatastrophic encoding circuit for Forney et al.'s
rate-1/3 quantum convolutional code \cite{isit2006grassl}. The basic
stabilizer and its first shift are as follows:%
\begin{equation}
\cdots\left\vert
\begin{array}
[c]{c}%
III\\
III\\
III\\
III
\end{array}
\right\vert
\begin{array}
[c]{c}%
XXX\\
ZZZ\\
III\\
III
\end{array}
\left\vert
\begin{array}
[c]{c}%
XZY\\
ZYX\\
XXX\\
ZZZ
\end{array}
\right\vert
\begin{array}
[c]{c}%
III\\
III\\
XZY\\
ZYX
\end{array}
\left\vert
\begin{array}
[c]{c}%
III\\
III\\
III\\
III
\end{array}
\right\vert \cdots\label{eq:qcc-example-stabilizer}%
\end{equation}
The code consists of all three-qubit shifts of the above generators. The
vertical bars are a visual aid to illustrate the three-qubit shifts of the
basic generators. The code can correct for an arbitrary single-qubit error in
every other frame.
\end{example}

\subsection{Stabilizer Entanglement Distillation without Entanglement
Assistance}

\label{sec:stabilizer-ent-distill}
\begin{figure}
[ptb]
\begin{center}
\includegraphics[
natheight=7.639800in,
natwidth=10.253200in,
height=2.2329in,
width=2.9914in
]%
{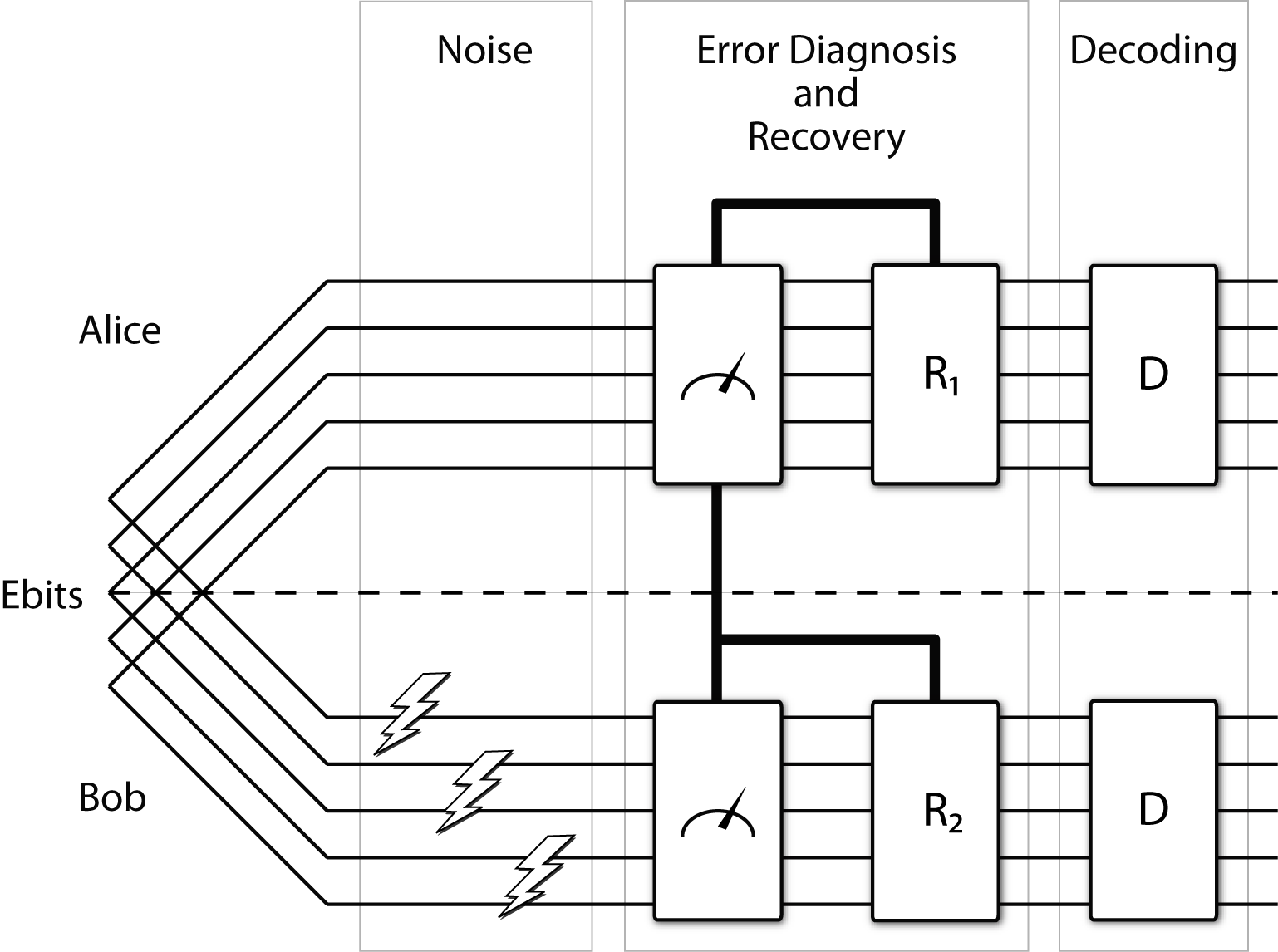}%
\caption{An example of a block entanglement distillation protocol. A sender
creates a set of noisy ebits by sending half of a set of Bell states through a
noisy quantum channel. Both sender and receiver perform multi-qubit
measurements to diagnose channel error. The sender transmits her measurement
results to the receiver over a classical communications channel. Both perform
recovery and decoding operations to obtain a set of noiseless ebits.}%
\label{fig:block-entanglement-distill}%
\end{center}
\end{figure}

The purpose of an $\left[  n,k\right]
$\ entanglement distillation protocol is to distill $k$ pure ebits from $n$
noisy ebits where $0\leq k\leq n$ \cite{PhysRevLett.76.722,PhysRevA.54.3824}.
The yield of such a protocol is $k/n$. Two parties can then use the noiseless
ebits for quantum communication protocols.
Figure~\ref{fig:block-entanglement-distill} illustrates the operation of a
block entanglement distillation protocol.%

The two parties establish a set of shared noisy ebits in the following way.
The sender Alice first prepares $n$ Bell states $\left\vert \Phi
^{+}\right\rangle ^{\otimes n}$ locally. She sends the second qubit of each
pair over a noisy quantum channel to a receiver Bob. Let $\left\vert \Phi
_{n}^{+}\right\rangle $ be the state $\left\vert \Phi^{+}\right\rangle
^{\otimes n}$ rearranged so that all of Alice's qubits are on the left and all
of Bob's qubits are on the right. The noisy channel applies a Pauli error in
the error set $\mathcal{E}\subset\Pi^{n}$ to the set of $n$ qubits sent over
the channel. The sender and receiver then share a set of $n$ noisy ebits of
the form $\left(  \mathbf{I}\otimes\mathbf{A}\right)  \left\vert \Phi_{n}%
^{+}\right\rangle $ where the identity $\mathbf{I}$ acts on Alice's qubits and
$\mathbf{A}$ is some Pauli operator in $\mathcal{E}$ acting on Bob's qubits.

A one-way stabilizer entanglement distillation protocol uses a stabilizer code
for the distillation procedure. Figure~\ref{fig:block-entanglement-distill}
highlights the main features of a stabilizer entanglement distillation
protocol. Suppose the stabilizer $\mathcal{S}$\ for an $\left[  n,k\right]
$\ quantum error-correcting code has generators $g_{1},\ldots,g_{n-k}$. The
distillation procedure begins with Alice measuring the $n-k$ generators in
$\mathcal{S}$. Let $\left\{  \mathbf{P}_{i}\right\}  $ be the set of the
$2^{n-k}$\ projectors that project onto the $2^{n-k}$ orthogonal subspaces
corresponding to the generators in $\mathcal{S}$. The measurement projects
$\left\vert \Phi_{n}^{+}\right\rangle $ randomly onto one of the
$i$\ subspaces. Each $\mathbf{P}_{i}$ commutes with the noisy operator
$\mathbf{A}$\ on Bob's side so that%
\begin{equation}
\left(  \mathbf{P}_{i}\otimes\mathbf{I}\right)  \left(  \mathbf{I}%
\otimes\mathbf{A}\right)  \left\vert \Phi_{n}^{+}\right\rangle =\left(
\mathbf{I}\otimes\mathbf{A}\right)  \left(  \mathbf{P}_{i}\otimes
\mathbf{I}\right)  \left\vert \Phi_{n}^{+}\right\rangle .
\label{eq:distill-proof}%
\end{equation}
The following important \textquotedblleft Bell-state\ matrix
identity\textquotedblright\ holds for an arbitrary matrix $\mathbf{M}$:%
\begin{equation}
\left(  \mathbf{M}\otimes\mathbf{I}\right)  \left\vert \Phi_{n}^{+}%
\right\rangle =\left(  \mathbf{I}\otimes\mathbf{M}^{T}\right)  \left\vert
\Phi_{n}^{+}\right\rangle .
\end{equation}
Then (\ref{eq:distill-proof}) is equal to the following:%
\begin{align}
\left(  \mathbf{I}\otimes\mathbf{A}\right)  \left(  \mathbf{P}_{i}%
\otimes\mathbf{I}\right)  \left\vert \Phi_{n}^{+}\right\rangle  &  =\left(
\mathbf{I}\otimes\mathbf{A}\right)  \left(  \mathbf{P}_{i}^{2}\otimes
\mathbf{I}\right)  \left\vert \Phi_{n}^{+}\right\rangle \nonumber\\
&  =\left(  \mathbf{I}\otimes\mathbf{A}\right)  \left(  \mathbf{P}_{i}%
\otimes\mathbf{P}_{i}^{T}\right)  \left\vert \Phi_{n}^{+}\right\rangle .
\end{align}
Therefore each of Alice's projectors $\mathbf{P}_{i}$ projects Bob's qubits
onto a subspace $\mathbf{P}_{i}^{T}$ corresponding to Alice's projected
subspace $\mathbf{P}_{i}$. Alice restores her qubits to the simultaneous
+1-eigenspace of the generators in $\mathcal{S}$. She sends her measurement
results to Bob. Bob measures the generators in $\mathcal{S}$. Bob combines his
measurements with Alice's to determine a syndrome for the error. He performs a
recovery operation on his qubits to reverse the error. He restores his qubits
to the simultaneous +1-eigenspace of the generators in $\mathcal{S}$. Alice
and Bob both perform the decoding unitary corresponding to stabilizer
$\mathcal{S}$\ to convert their $k$ logical ebits to $k$ physical ebits.

\subsection{Stabilizer Entanglement Distillation with Entanglement Assistance}

\label{sec:stabilizer-ent-assist-ent-distill}Luo and Devetak provided a
straightforward extension of the above protocol \cite{luo:010303}. Their
method converts an entanglement-assisted stabilizer code into an
entanglement-assisted entanglement distillation protocol.

Luo and Devetak form an entanglement distillation protocol that has
entanglement assistance from a few noiseless ebits. The crucial assumption for
an entanglement-assisted entanglement distillation protocol is that Alice and
Bob possess $c$ noiseless ebits in addition to their $n$ noisy ebits. The
total state of the noisy and noiseless ebits is%
\begin{equation}
(\mathbf{I}^{A}\otimes\left(  \mathbf{A\otimes I}\right)  ^{B})\left\vert
\Phi_{n+c}^{+}\right\rangle
\end{equation}
where $\mathbf{I}^{A}$ is the $2^{n+c}\times2^{n+c}$ identity matrix acting on
Alice's qubits and the noisy Pauli operator $\left(  \mathbf{A\otimes
I}\right)  ^{B}$ affects Bob's first $n$ qubits only. Thus the last $c$ ebits
are noiseless, and Alice and Bob have to correct for errors on the first $n$
ebits only.

The protocol proceeds exactly as outlined in the previous section. The only
difference is that Alice and Bob measure the generators in an
entanglement-assisted stabilizer code. Each generator spans over $n+c$ qubits
where the last $c$ qubits are noiseless.

We comment on the yield of this entanglement-assisted entanglement
distillation protocol. An entanglement-assisted code has $n-k$ generators that
each have $n+c$ Pauli entries. These parameters imply that the entanglement
distillation protocol produces $k+c$ ebits. But the protocol consumes $c$
initial noiseless ebits as a catalyst for distillation. Therefore the yield of
this protocol is $k/n$.

In Section \ref{sec:conv-ent-ent-assist}, we exploit this same idea of using a
few noiseless ebits as a catalyst for distillation. The idea is similar in
spirit to that developed in this section, but the mathematics and construction
are different because we perform distillation in a convolutional manner.

\section{Convolutional Entanglement Distillation without Entanglement
Assistance}%

\begin{figure*}
[ptb]
\begin{center}
\includegraphics[
natheight=8.013400in,
natwidth=19.253300in,
height=2.8807in,
width=6.8978in
]
{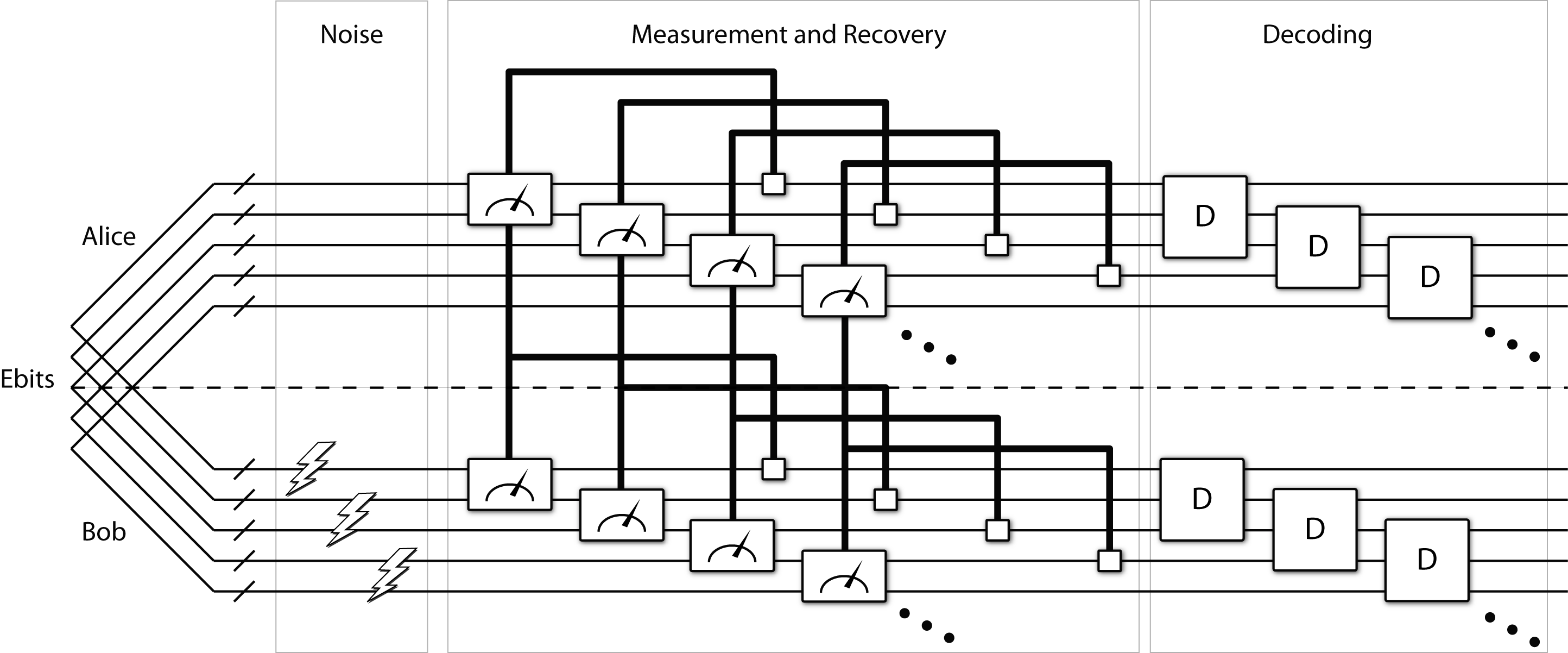}
\caption
{An example of a convolutional entanglement distillation protocol taken
from the quantum convolutional code in Ref. \cite{ieee2007forney}.
The code in Ref. \cite{ieee2007forney} has rate 1/3 and can correct
for single-qubit errors in every other frame. Alice and Bob first measure the
operators in the stabilizer for the quantum convolutional code. Alice
performs conditional unitaries on her qubits to restore them
to the +1 eigenspace of the stabilizer code. Alice forwards her measurement results to Bob. Bob performs a
maximum-likelihood decoding procedure such as Viterbi decoding \cite
{itit1967viterbi} to determine
the qubit errors. He corrects for these errors. He restores his qubits to the +1 eigenspace of the stabilizer code.
Alice and Bob both perform
online decoding to obtain ebits with yield 1/3.}
\label{fig:cedc}
\end{center}
\end{figure*}%
\label{sec:conv-ent-dist}We now show how to convert an arbitrary quantum
convolutional code into a convolutional entanglement distillation protocol.
Figure~\ref{fig:cedc}\ illustrates an example of a yield-1/3 convolutional
entanglement distillation protocol. The protocol has the same benefits as a
quantum convolutional code:\ an online decoder with less decoding complexity
than a block protocol, good error-correcting properties, and higher ebit yield
than a block protocol. The protocol we develop in this section is useful for
our major contribution presented in the next section.

We can think of our protocol in two ways. Our protocol applies when a sender
Alice and a receiver Bob possess a countably infinite number of noisy ebits.
Our protocol also applies as an online protocol when Alice and Bob begin with
a finite number of noisy ebits and establish more as time passes. The
countably infinite and online protocols are equivalent. We would actually
implement the entanglement distillation protocol in the online manner, but we
formulate the forthcoming mathematics with the countably infinite description.
Each step in the protocol does not need to wait for the completion of its
preceding step if Alice and Bob employ the protocol online.

The protocol begins with Alice and Bob establishing a set of noisy ebits.
Alice prepares a countably infinite number of Bell states $\left\vert \Phi
^{+}\right\rangle $ locally. She sends one half of each Bell state through a
noisy quantum channel. Alice and Bob then possess a state $\rho^{AB}$\ that is
a countably infinite number of noisy ebits $\rho_{i}^{AB}$ where%
\begin{equation}
\rho^{AB}=%
{\displaystyle\bigotimes\limits_{i=1}^{\infty}}
\ \rho_{i}^{AB}.
\end{equation}
The state $\rho^{AB}$ is equivalent to the following ensemble%
\begin{equation}
\left\{  p_{i},\left\vert \Phi^{+}\right\rangle _{i}^{AB}\right\}  .
\end{equation}
In the above, $p_{i}$ is the probability that the state is $\left\vert
\Phi^{+}\right\rangle _{i}^{AB}$,%
\begin{equation}
\left\vert \Phi^{+}\right\rangle _{i}^{AB}\equiv\left(  \mathbf{I}%
\otimes\mathbf{A}_{i}\right)  \left\vert \Phi_{\infty}^{+}\right\rangle ^{AB},
\end{equation}
and $\left\vert \Phi_{\infty}^{+}\right\rangle ^{AB}$ is the state $\left(
\left\vert \Phi^{+}\right\rangle ^{AB}\right)  ^{\otimes\infty}$ rearranged so
that all of Alice's qubits are on the left and all of Bob's are on the right.
$\mathbf{A}_{i}\in\Pi^{\mathbb{Z}^{+}}$ is a Pauli sequence of errors acting
on Bob's side. These errors result from the noisy quantum channel.
$\mathbf{I}$ is a sequence of identity matrices acting on Alice's side
indicating that the noisy channel does not affect her qubits. Alice and Bob
need to correct for a particular error set in order to distill noiseless ebits.

Alice and Bob employ the following strategy to distill noiseless ebits. Alice
measures the $n-k$ generators in the basic set $\mathcal{G}_{0}$. The
measurement operation projects the first $n\left(  \nu+1\right)  $ ebits
($\nu$ is the constraint length) randomly onto one of $2^{n-k}$\ orthogonal
subspaces. Alice places the measurement outcomes in an $\left(  n-k\right)
$-dimensional classical bit vector $\mathbf{a}_{0}$. She restores her half of
the noisy ebits to the simultaneous +1-eigenspace of the generators in
$\mathcal{G}_{0}$ if $\mathbf{a}_{0}$ differs from the all-zero vector. She
sends $\mathbf{a}_{0}$ to Bob over a classical communications channel. Bob
measures the generators in $\mathcal{G}_{0}$ and stores the measurement
outcomes in a classical bit vector $\mathbf{b}_{0}$. Bob compares
$\mathbf{b}_{0}$ to $\mathbf{a}_{0}$ by calculating an error vector
$\mathbf{e}_{0}=\mathbf{a}_{0}\oplus\mathbf{b}_{0}$. He corrects for any
errors that $\mathbf{e}_{0}$ can identify. He may have to wait to receive
later error vectors before determining the full error syndrome. He restores
his half of the noisy ebits to the simultaneous +1-eigenspace of the
generators in $\mathcal{G}_{0}$ if the bit vector $\mathbf{b}_{0}$ indicates
that his logical ebits are not in the +1-space. Alice and Bob repeat the above
procedure for all shifts $D\left(  \mathcal{G}_{0}\right)  $, $D^{2}\left(
\mathcal{G}_{0}\right)  $, \ldots\ of the basic generators in $\mathcal{G}%
_{0}$. Bob obtains a set $\mathcal{E}$ of classical error vectors
$\mathbf{e}_{i}$: $\mathcal{E}=\left\{  \mathbf{e}_{i}:i\in\mathbb{Z}%
^{+}\right\}  $. Bob uses a maximum-likelihood decoding technique such as
Viterbi decoding \cite{itit1967viterbi} or a table-lookup on the error set
$\mathcal{E}$ to determine which errors occur. This error determination
process is a purely classical computation. He reverses all errors after
determining the syndrome.

The states that Alice and Bob possess after the above procedure are encoded
logical ebits. They can extract physical ebits from these logical ebits by
each performing the online decoding circuit for the code $\mathcal{G}$. The
algorithm outlined in \cite{isit2006grassl}\ gives a method for determining
the online decoding circuit.

\begin{example}
We use the rate-1/3 quantum convolutional code in
Example~\ref{sec:qcc-example} to produce a yield-1/3 convolutional
entanglement distillation protocol. Alice measures the generators in the
stabilizer in (\ref{eq:qcc-example-stabilizer})\ for every noisy ebit she
shares with Bob. Alice communicates the result of her measurement of the first
two generators to Bob. Alice restores the qubits on her side to be in the
simultaneous +1-eigenspace of the first two generators. Bob measures the same
first two generators. Alice measures the next two generators, communicates her
results, etc. Bob compares his results to Alice's to determine the error bit
vectors. Bob performs Viterbi decoding on the measurement results and corrects
for errors. He rotates his states to the simultaneous +1-eigenspace of the
generators. Alice and Bob perform the above procedure in an online manner
according to Figure~\ref{fig:cedc}. Alice and Bob can decode the first six
qubits after measuring the second two generators. They can decode because
there is no overlap between the first two generators and any two generators
after the second two generators. They use the circuit from
\cite{isit2006grassl} in reverse order to decode physical ebits from logical
ebits. They distill ebits with yield 1/3 by using this convolutional
entanglement distillation protocol. The ebit yield of 1/3 follows directly
from the code rate of 1/3.
\end{example}

\section{Convolutional Entanglement Distillation with Entanglement Assistance}

\label{sec:conv-ent-ent-assist}The convolutional entanglement distillation
protocol that we develop in this section operates identically to the one
developed in the previous section. The measurements, classical communication,
and recovery and decoding operations proceed exactly as Figure~\ref{fig:cedc} indicates.

The difference between the protocol in this section and the previous one is
that we now assume the sender and receiver share a few initial noiseless
ebits. They use these initial ebits as a catalyst to get the protocol started.
The sender and receiver require noiseless ebits for each round of the
convolutional entanglement distillation protocol. They can use the noiseless
ebits generated by earlier rounds for consumption in later rounds. It is
possible to distill noiseless ebits in this way by catalyzing the process with
a few noiseless ebits. The protocol we develop in this section is a more
powerful generalization of the previous section's protocol.

The construction in this section allows sender and receiver to use an
arbitrary set of Paulis for the distillation protocol. The set does not
necessarily have to be a commuting set of Paulis. The idea is similar in
spirit to entanglement-assisted quantum error correction
\cite{science2006brun,arx2006brun}.

The implication of the construction in this section is that we can import an
arbitrary binary or quaternary classical convolutional code for use as a
quantum convolutional code. We explicitly give some examples to highlight the
technique for importing. The error-correcting properties and yield translate
directly from the properties of the classical convolutional code. Thus the
problem of finding a good convolutional entanglement distillation protocol
reduces to that of finding a good classical convolutional code.

We first review some mathematics concerning the commutative properties of
quantum convolutional codes. We then present our different constructions for a
convolutional entanglement distillation protocol that use entanglement assistance.

\subsection{Commutative Properties of Quantum Convolutional Codes}

We consider the commutative properties of quantum convolutional codes. We
develop some mathematics that leads to the important \textquotedblleft shifted
symplectic product.\textquotedblright\ The shifted symplectic product reveals
the commutation relations of an arbitrary number of shifts of a set of Pauli
sequences. All of our constructions following this preliminary section exploit
the properties of the shifted symplectic product.

We first define the phase-free Pauli group $\left[  \Pi^{\mathbb{Z}}\right]
$\ on a sequence of qubits. Recall that the delay transform $D$ in
(\ref{eq:delay-transform}) shifts a Pauli sequence to the right by $n$. Let us
assume for now that $n=1$. Let $\Pi^{\mathbb{Z}}$ denote the set of all
countably infinite Pauli sequences. The set $\Pi^{\mathbb{Z}}$ is equivalent
to the set of all one-qubit shifts of arbitrary Pauli operators:%
\begin{equation}
\Pi^{\mathbb{Z}}=\left\{
{\textstyle\prod\limits_{i\in\mathbb{Z}}}
D^{i}\left(  A_{i}\right)  :A_{i}\in\Pi\right\}  .
\end{equation}
We remark that $D^{i}\left(  A_{i}\right)  =D^{i}\left(  A_{i}\otimes
I^{\otimes\infty}\right)  $. We make this same abuse of notation in what
follows. We can define the equivalence class $\left[  \Pi^{\mathbb{Z}}\right]
$\ of phase-free Pauli sequences:%
\begin{equation}
\left[  \Pi^{\mathbb{Z}}\right]  =\left\{  \beta\mathbf{A\ }|\ \mathbf{A}%
\in\Pi^{\mathbb{Z}},\beta\in\mathbb{C},\left\vert \beta\right\vert =1\right\}
.
\end{equation}

We develop a relation between binary polynomials and Pauli sequences that is
useful for representing the shifting nature of quantum convolutional codes.
Suppose $z\left(  D\right)  $ and $x\left(  D\right)  $ are arbitrary
finite-degree and finite-delay polynomials in $D$\ over $\mathbb{Z}_{2}$%
\begin{align}
z\left(  D\right)   &  =\sum_{i}z_{i}D^{i},\ \ \ \ \ \ \ z_{i}\in
\mathbb{Z}_{2}\ \ \forall i\in\mathbb{Z},\\
x\left(  D\right)   &  =\sum_{i}x_{i}D^{i},\ \ \ \ \ \ \ x_{i}\in
\mathbb{Z}_{2}\ \ \forall i\in\mathbb{Z},
\end{align}
where del$\left(  z\left(  D\right)  \right)  $, del$\left(  x\left(
D\right)  \right)  $, deg$\left(  z\left(  D\right)  \right)  $, deg$\left(
x\left(  D\right)  \right)  <\infty$. Suppose%
\begin{equation}
u\left(  D\right)  =\left(  z\left(  D\right)  ,x\left(  D\right)  \right)
\in\left(  \mathbb{Z}_{2}\left(  D\right)  \right)  ^{2},
\end{equation}
where $\left(  \mathbb{Z}_{2}\left(  D\right)  \right)  ^{2}$ indicates the
direct product $\mathbb{Z}_{2}\left(  D\right)  \times\mathbb{Z}_{2}\left(
D\right)  $. Let us employ the following shorthand:%
\begin{equation}
u\left(  D\right)  =\left(  z\left(  D\right)  |x\left(  D\right)  \right)  .
\end{equation}
Let $N$ be a map from the binary polynomials to the Pauli sequences,
$N:\left(  \mathbb{Z}_{2}\left(  D\right)  \right)  ^{2}\rightarrow
\Pi^{\mathbb{Z}}$, where%
\begin{equation}
N\left(  u\left(  D\right)  \right)  =%
{\textstyle\prod\limits_{i}}
D^{i}\left(  Z^{z_{i}}X^{x_{i}}\right)  .
\end{equation}
Let $v\left(  D\right)  =\left(  z^{\prime}\left(  D\right)  |x^{\prime
}\left(  D\right)  \right)  $ where $v\left(  D\right)  \in\left(
\mathbb{Z}_{2}\left(  D\right)  \right)  ^{2}$. The map $N$ induces an
isomorphism%
\begin{equation}
\left[  N\right]  :\left(  \mathbb{Z}_{2}\left(  D\right)  \right)
^{2}\rightarrow\left[  \Pi^{\mathbb{Z}}\right]  ,
\end{equation}
because addition of binary polynomials is equivalent to multiplication of
Pauli elements up to a global phase:%
\begin{equation}
\left[  N\left(  u\left(  D\right)  +v\left(  D\right)  \right)  \right]
=\left[  N\left(  u\left(  D\right)  \right)  \right]  \left[  N\left(
v\left(  D\right)  \right)  \right]  . \label{eq:isomorphism-poly}%
\end{equation}
The above isomorphism is a powerful way to capture the infiniteness and
shifting nature of convolutional codes with finite-degree and finite-delay
polynomials over the binary field $\mathbb{Z}_{2}$.

Recall from Definition~\ref{def:conv-code} that a commuting set comprising a
basic set of Paulis and all their shifts specifies a quantum convolutional
code. How can we capture the commutation relations of a Pauli sequence and all
of its shifts?\ The \textit{shifted} symplectic product $\odot$, where%
\begin{equation}
\odot:\left(  \mathbb{Z}_{2}\left(  D\right)  \right)  ^{2}\times\left(
\mathbb{Z}_{2}\left(  D\right)  \right)  ^{2}\rightarrow\mathbb{Z}_{2}\left(
D\right)  ,
\end{equation}
is an elegant way to do so. The shifted symplectic product maps two vectors
$u\left(  D\right)  $ and $v\left(  D\right)  $ to a binary polynomial with
finite delay and finite degree:%
\begin{equation}
\left(  u\odot v\right)  \left(  D\right)  =z\left(  D^{-1}\right)  x^{\prime
}\left(  D\right)  -x\left(  D^{-1}\right)  z^{\prime}\left(  D\right)  .
\end{equation}
The symplectic orthogonality condition originally given in Ref.
\cite{arxiv2004olliv} inspires the definition for the shifted symplectic
product. The shifted symplectic product is not a proper symplectic product
because it fails to be alternating \cite{book2001symp}. The alternating
property requires that%
\begin{equation}
\left(  u\odot v\right)  \left(  D\right)  =-\left(  v\odot u\right)  \left(
D\right)  ,
\end{equation}
but we find instead that the following holds:%
\begin{equation}
\left(  u\odot v\right)  \left(  D\right)  =-\left(  v\odot u\right)  \left(
D^{-1}\right)  .
\end{equation}
Every vector $u\left(  D\right)  \in\mathbb{Z}_{2}\left(  D\right)  ^{2}$ is
self-time-reversal antisymmetric with respect to $\odot$:%
\begin{equation}
\left(  u\odot u\right)  \left(  D\right)  =-\left(  u\odot u\right)  \left(
D^{-1}\right)  \ \ \ \ \forall u\left(  D\right)  \in\mathbb{Z}_{2}\left(
D\right)  ^{2}.
\end{equation}
Every binary vector is also self-time-reversal symmetric with respect to
$\odot$\ because addition and subtraction are the same over $\mathbb{Z}_{2}$.
We employ the addition convention from now on and drop the minus signs. The
shifted symplectic product is a binary polynomial in $D$. We write its
coefficients as follows:%
\begin{equation}
\left(  u\odot v\right)  \left(  D\right)  =\sum_{i\in\mathbb{Z}}\left(
u\odot v\right)  _{i}\ D^{i}.
\end{equation}
The coefficient $\left(  u\odot v\right)  _{i}$ captures the commutation
relations of two Pauli sequences for $i$ $n$-qubit shifts of one of the
sequences:%
\begin{multline}
N\left(  u\left(  D\right)  \right)  D^{i}\left(  N\left(  v\left(  D\right)
\right)  \right)  =\label{eq:shifted-symp-comm}\\
\left(  -1\right)  ^{\left(  u\odot v\right)  _{i}}D^{i}\left(  N\left(
v\left(  D\right)  \right)  \right)  N\left(  u\left(  D\right)  \right)  .
\end{multline}
Thus two Pauli sequences $N\left(  u\left(  D\right)  \right)  $ and $N\left(
v\left(  D\right)  \right)  $\ commute for all shifts if and only if the
shifted symplectic product $\left(  u\odot v\right)  \left(  D\right)  $ vanishes.

The next example highlights the main features of the shifted symplectic
product and further emphasizes the relationship between Pauli commutation and
orthogonality of the shifted symplectic product.

\begin{example}
Consider two sets of binary polynomials:%
\begin{align}
z_{1}\left(  D\right)   &  =D,\ \ \ \ x_{1}\left(  D\right)  =1+D^{3}%
,\nonumber\\
z_{2}\left(  D\right)   &  =1+D,\ \ \ \ x_{2}\left(  D\right)  =D^{3}%
.\nonumber
\end{align}
We form vectors $u\left(  D\right)  $ and $v\left(  D\right)  $ from the above
polynomials where%
\begin{align}
u\left(  D\right)   &  =\left(  z_{1}\left(  D\right)  \ |\ x_{1}\left(
D\right)  \right)  ,\nonumber\\
v\left(  D\right)   &  =\left(  z_{2}\left(  D\right)  \ |\ x_{2}\left(
D\right)  \right)  .
\end{align}
The isomorphism $N$ maps the above polynomials to the following Pauli
sequences:%
\begin{equation}%
\begin{array}
[c]{c}%
N\left(  u\left(  D\right)  \right)  =\left(  \cdots|I|X|Z|I|X|I|\cdots
\right)  ,\\
N\left(  v\left(  D\right)  \right)  =\left(  \cdots|I|Z|Z|I|X|I|\cdots
\right)  .
\end{array}
\end{equation}
The vertical bars between every Pauli in the sequence indicate that we are
considering one-qubit shifts. We determine the commutation relations of the
above sequences by inspection. $N\left(  u\left(  D\right)  \right)  $
anticommutes with a shift of itself by one or two to the left or right and
commutes with all other shifts of itself. $N\left(  v\left(  D\right)
\right)  $ anticommutes with a shift of itself by two or three to the left or
right and commutes with all other shifts of itself. $N\left(  u\left(
D\right)  \right)  $ anticommutes with $N\left(  v\left(  D\right)  \right)  $
shifted to the left by one or two, with the zero-shifted $N\left(  v\left(
D\right)  \right)  $, and with $N\left(  v\left(  D\right)  \right)  $ shifted
to the right by two or three. The following shifted symplectic products give
us the same information:%
\begin{align}
\left(  u\odot u\right)  \left(  D\right)   &  =D^{-2}+D^{-1}+D+D^{2}%
,\nonumber\\
\left(  v\odot v\right)  \left(  D\right)   &  =D^{-3}+D^{-2}+D^{2}%
+D^{3},\nonumber\\
\left(  v\odot u\right)  \left(  D\right)   &  =D^{-2}+D^{-1}+1+D^{2}+D^{3}.
\end{align}
The nonzero coefficients indicate the commutation relations just as
(\ref{eq:shifted-symp-comm}) claims.
\end{example}

A quantum convolutional code in general consists of generators with $n$ qubits
per frame. Therefore, we consider the $n$-qubit extension of the definitions
and isomorphism given above. Let the delay transform $D$ now shift a Pauli
sequence to the right by an arbitrary integer $n$. Consider a $2n$-dimensional
vector $\mathbf{u}\left(  D\right)  $ of binary polynomials where
$\mathbf{u}\left(  D\right)  \in\left(  \mathbb{Z}_{2}\left(  D\right)
\right)  ^{2n}$. Let us write $\mathbf{u}\left(  D\right)  $\ as follows%
\begin{align*}
\mathbf{u}\left(  D\right)   &  =\left(  \mathbf{z}\left(  D\right)
|\mathbf{x}\left(  D\right)  \right)  ,\\
&  =\left(
\begin{array}
[c]{ccc}%
z_{1}\left(  D\right)  & \cdots & z_{n}\left(  D\right)
\end{array}
|%
\begin{array}
[c]{ccc}%
x_{1}\left(  D\right)  & \cdots & x_{n}\left(  D\right)
\end{array}
\right)  ,
\end{align*}
where $\mathbf{z}\left(  D\right)  ,\mathbf{x}\left(  D\right)  \in\left(
\mathbb{Z}_{2}\left(  D\right)  \right)  ^{n}$. Suppose%
\begin{align}
z_{i}\left(  D\right)   &  =\sum_{j}z_{i,j}D^{j},\nonumber\\
x_{i}\left(  D\right)   &  =\sum_{j}x_{i,j}D^{j}.
\end{align}
Define a map $\mathbf{N}:\left(  \mathbb{Z}_{2}\left(  D\right)  \right)
^{2n}\rightarrow\Pi^{\mathbb{Z}}$:%
\begin{multline}
\mathbf{N}\left(  \mathbf{u}\left(  D\right)  \right)  =%
{\textstyle\prod\limits_{j}}
D^{j}\left(  Z^{z_{1,j}}X^{x_{1,j}}\right) \nonumber\\
D^{j}\left(  I\otimes Z^{z_{2,j}}X^{x_{2,j}}\right)  \cdots D^{j}\left(
I^{\otimes n-1}\otimes Z^{z_{n,j}}X^{x_{n,j}}\right)  .
\end{multline}
$\mathbf{N}$ is equivalent to the following map (up to a global phase)%
\begin{multline}
\mathbf{N}\left(  \mathbf{u}\left(  D\right)  \right)  =N\left(  u_{1}\left(
D\right)  \right)  \left(  I\otimes N\left(  u_{2}\left(  D\right)  \right)
\right) \nonumber\\
\cdots\left(  I^{\otimes n-1}\otimes N\left(  u_{n}\left(  D\right)  \right)
\right)  ,
\end{multline}
where%
\begin{equation}
u_{i}\left(  D\right)  =\left(  z_{i}\left(  D\right)  |x_{i}\left(  D\right)
\right)  .
\end{equation}
Suppose%
\begin{equation}
\mathbf{v}\left(  D\right)  =\left(  \mathbf{z}^{\prime}\left(  D\right)
|\mathbf{x}^{\prime}\left(  D\right)  \right)  ,
\end{equation}
where $\mathbf{v}\left(  D\right)  \in\left(  \mathbb{Z}_{2}\left(  D\right)
\right)  ^{2n}$. The map $\mathbf{N}$\ induces an isomorphism $\left[
\mathbf{N}\right]  :\left(  \mathbb{Z}_{2}\left(  D\right)  \right)
^{2n}\rightarrow\left[  \Pi^{\mathbb{Z}}\right]  $ for the same reasons given
in (\ref{eq:isomorphism-poly}):%
\begin{equation}
\left[  \mathbf{N}\left(  \mathbf{u}\left(  D\right)  +\mathbf{v}\left(
D\right)  \right)  \right]  =\left[  \mathbf{N}\left(  \mathbf{u}\left(
D\right)  \right)  \right]  \left[  \mathbf{N}\left(  \mathbf{v}\left(
D\right)  \right)  \right]  .
\end{equation}
The isomorphism $\mathbf{N}$ is again useful because it allows us to perform
binary calculations instead of Pauli calculations.

We can again define a shifted symplectic product for the case of $n$-qubits
per frame. Let $\odot$ denote the shifted symplectic product between vectors
of binary polynomials:%
\begin{equation}
\odot:\left(  \mathbb{Z}_{2}\left(  D\right)  \right)  ^{2n}\times\left(
\mathbb{Z}_{2}\left(  D\right)  \right)  ^{2n}\rightarrow\mathbb{Z}_{2}\left(
D\right)  .
\end{equation}
It maps vectors of binary polynomials to a finite-degree and finite-delay
binary polynomial%
\begin{equation}
\left(  \mathbf{u}\odot\mathbf{v}\right)  \left(  D\right)  =\sum_{i=1}%
^{n}\left(  u_{i}\odot v_{i}\right)  \left(  D\right)  ,
\end{equation}
where%
\begin{align*}
u_{i}\left(  D\right)   &  =\left(  z_{i}\left(  D\right)  |x_{i}\left(
D\right)  \right)  ,\\
v_{i}\left(  D\right)   &  =\left(  z_{i}^{\prime}\left(  D\right)
|x_{i}^{\prime}\left(  D\right)  \right)  .
\end{align*}
The standard inner product gives an alternative way to define the shifted
symplectic product:%
\begin{equation}
\left(  \mathbf{u}\odot\mathbf{v}\right)  \left(  D\right)  =\mathbf{z}\left(
D^{-1}\right)  \cdot\mathbf{x}^{\prime}\left(  D\right)  -\mathbf{x}\left(
D^{-1}\right)  \cdot\mathbf{z}^{\prime}\left(  D\right)  .
\end{equation}
Every vector $\mathbf{u}\left(  D\right)  \in\mathbb{Z}_{2}\left(  D\right)
^{2n}$ is self-time-reversal symmetric with respect to $\odot$:%
\begin{equation}
\left(  \mathbf{u}\odot\mathbf{u}\right)  \left(  D\right)  =\left(
\mathbf{u}\odot\mathbf{u}\right)  \left(  D^{-1}\right)  \ \ \ \ \forall
\mathbf{u}\left(  D\right)  \in\mathbb{Z}_{2}\left(  D\right)  ^{2n}.
\label{eq:self-time-reversal-sym}%
\end{equation}
The shifted symplectic product for vectors of binary polynomials is a binary
polynomial in $D$. We write its coefficients as follows:%
\begin{equation}
\left(  \mathbf{u}\odot\mathbf{v}\right)  \left(  D\right)  =\sum
_{i\in\mathbb{Z}}\left(  \mathbf{u}\odot\mathbf{v}\right)  _{i}\ D^{i}.
\end{equation}
The coefficient $\left(  \mathbf{u}\odot\mathbf{v}\right)  _{i}$\ captures the
commutation relations of two Pauli sequences for $i$ $n$-qubit shifts of one
of the sequences:%
\begin{multline}
\mathbf{N}\left(  \mathbf{u}\left(  D\right)  \right)  D^{i}\left(
\mathbf{N}\left(  \mathbf{v}\left(  D\right)  \right)  \right)  =\nonumber\\
\left(  -1\right)  ^{\left(  \mathbf{u}\odot\mathbf{v}\right)  _{i}}%
D^{i}\left(  \mathbf{N}\left(  \mathbf{v}\left(  D\right)  \right)  \right)
\mathbf{N}\left(  \mathbf{u}\left(  D\right)  \right)  .
\end{multline}

\begin{example}
We consider the case where $n=4$. Consider the following vectors of
polynomials:%
\begin{equation}
\left[
\begin{array}
[c]{c}%
\mathbf{z}\left(  D\right) \\
\mathbf{x}\left(  D\right) \\
\mathbf{z}^{\prime}\left(  D\right) \\
\mathbf{x}^{\prime}\left(  D\right)
\end{array}
\right]  =\left[
\begin{array}
[c]{cccc}%
1+D & D & 1 & D\\
0 & 1 & 0 & 0\\
0 & 1 & 0 & 0\\
1+D & 1+D & 1 & D
\end{array}
\right]  .
\end{equation}
Suppose%
\begin{align}
\mathbf{u}\left(  D\right)   &  =\left(  \mathbf{z}\left(  D\right)
|\mathbf{x}\left(  D\right)  \right)  ,\nonumber\\
\mathbf{v}\left(  D\right)   &  =\left(  \mathbf{z}^{\prime}\left(  D\right)
|\mathbf{x}^{\prime}\left(  D\right)  \right)  .
\end{align}
The isomorphism $\mathbf{N}$ maps $\mathbf{u}\left(  D\right)  $ and
$\mathbf{v}\left(  D\right)  $ to the following Pauli sequences:%
\begin{align}
\mathbf{N}\left(  \mathbf{u}\left(  D\right)  \right)   &  =\left(
\cdots|IIII|ZXZI|ZZIZ|IIII|\cdots\right)  ,\nonumber\\
\mathbf{N}\left(  \mathbf{v}\left(  D\right)  \right)   &  =\left(
\cdots|IIII|XYXI|XXIX|IIII|\cdots\right)  .
\end{align}
We can determine the commutation relations by inspection of the above Pauli
sequences. $\mathbf{N}\left(  \mathbf{u}\left(  D\right)  \right)  $
anticommutes with itself shifted by one to the left or right, $\mathbf{N}%
\left(  \mathbf{v}\left(  D\right)  \right)  $ anticommutes with itself
shifted by one to the left or right, and $\mathbf{N}\left(  \mathbf{u}\left(
D\right)  \right)  $ anticommutes with $\mathbf{N}\left(  \mathbf{v}\left(
D\right)  \right)  $ shifted by one to the left. The following shifted
symplectic products confirm the above commutation relations:%
\begin{align}
\left(  \mathbf{u}\odot\mathbf{u}\right)  \left(  D\right)   &  =D^{-1}%
+D,\nonumber\\
\left(  \mathbf{v}\odot\mathbf{v}\right)  \left(  D\right)   &  =D^{-1}%
+D,\nonumber\\
\left(  \mathbf{u}\odot\mathbf{v}\right)  \left(  D\right)   &  =D.
\end{align}

\end{example}

We note two useful properties of the shifted symplectic product $\odot$.
Suppose $f\left(  D\right)  \in\mathbb{Z}_{2}\left(  D\right)  $ with
$\deg\left(  f\right)  \geq0$. Let us denote scalar polynomial multiplication
as follows:%
\begin{equation}
\left(  f\ \mathbf{u}\right)  \left(  D\right)  =%
\begin{bmatrix}
f\left(  D\right)  u_{1}\left(  D\right)  & \cdots & f\left(  D\right)
u_{n}\left(  D\right)
\end{bmatrix}
.
\end{equation}
The following identities hold.%
\begin{align}
\left(  \left(  f\ \mathbf{u}\right)  \odot\mathbf{v}\right)  \left(
D\right)   &  =f\left(  D^{-1}\right)  \left(  \mathbf{u}\odot\mathbf{v}%
\right)  \left(  D\right)  ,\\
\left(  \mathbf{u}\odot\left(  f\ \mathbf{v}\right)  \right)  \left(
D\right)   &  =f\left(  D\right)  \left(  \mathbf{u}\odot\mathbf{v}\right)
\left(  D\right)  .
\end{align}
We also remark that
\[
\left(  \mathbf{u}\odot\mathbf{v}\right)  \left(  D\right)  =\left(
\mathbf{v}\odot\mathbf{u}\right)  \left(  D\right)  ,
\]
iff%
\[
\left(  \mathbf{u}\odot\mathbf{v}\right)  \left(  D\right)  =\left(
\mathbf{u}\odot\mathbf{v}\right)  \left(  D^{-1}\right)  .
\]
We exploit both of the above properties in the constructions that follow.

\subsection{Yield (n-1)/n Convolutional Entanglement Distillation with
Entanglement Assistance}

We present our first method for constructing a convolutional entanglement
distillation protocol that uses entanglement assistance. The shifted
symplectic product is a crucial component of our formulation.

Suppose Alice and Bob use one generator $\mathbf{N}\left(  \mathbf{u}\left(
D\right)  \right)  $\ for an entanglement distillation protocol where%
\begin{align*}
\mathbf{u}\left(  D\right)   &  =\left(  \mathbf{z}\left(  D\right)
|\mathbf{x}\left(  D\right)  \right) \\
&  =\left(
\begin{array}
[c]{ccc}%
z_{1}\left(  D\right)  & \cdots & z_{n}\left(  D\right)
\end{array}
|%
\begin{array}
[c]{ccc}%
x_{1}\left(  D\right)  & \cdots & x_{n}\left(  D\right)
\end{array}
\right)  .
\end{align*}
We do not impose a commuting constraint on generator $\mathbf{N}\left(
\mathbf{u}\left(  D\right)  \right)  $. Alice and Bob choose generator
$\mathbf{N}\left(  \mathbf{u}\left(  D\right)  \right)  $ solely for its
error-correcting capability.

The shifted symplectic product helps to produce a commuting generator from a
noncommuting one. The shifted symplectic product of $\mathbf{u}\left(
D\right)  $\ is%
\begin{equation}
\left(  \mathbf{u}\odot\mathbf{u}\right)  \left(  D\right)  =\sum
_{i\in\mathbb{Z}}\left(  \mathbf{u}\odot\mathbf{u}\right)  _{i}\ D^{i}.
\end{equation}
The coefficient $\left(  \mathbf{u}\odot\mathbf{u}\right)  _{0}$ for zero
shifts is equal to zero because every tensor product of Pauli operators
commutes with itself:%
\begin{equation}
\left(  \mathbf{u}\odot\mathbf{u}\right)  _{0}=0.
\end{equation}
Recall that $\mathbf{u}\left(  D\right)  $ is self-time-reversal symmetric
(\ref{eq:self-time-reversal-sym}). We adopt the following notation for a
polynomial that includes the positive-index or negative-index coefficients of
the shifted symplectic product $\left(  \mathbf{u}\odot\mathbf{u}\right)
\left(  D\right)  $:%
\begin{align}
\left(  \mathbf{u}\odot\mathbf{u}\right)  \left(  D\right)  ^{+}  &
=\sum_{i\in\mathbb{Z}^{+}}\left(  \mathbf{u}\odot\mathbf{u}\right)
_{i}\ D^{i},\\
\left(  \mathbf{u}\odot\mathbf{u}\right)  \left(  D\right)  ^{-}  &
=\sum_{i\in\mathbb{Z}^{-}}\left(  \mathbf{u}\odot\mathbf{u}\right)
_{i}\ D^{i}.
\end{align}
The following identity holds:%
\begin{equation}
\left(  \mathbf{u}\odot\mathbf{u}\right)  \left(  D\right)  ^{+}=\left(
\mathbf{u}\odot\mathbf{u}\right)  \left(  D^{-1}\right)  ^{-}.
\end{equation}
Consider the following vector of polynomials:%
\begin{equation}
\mathbf{a}\left(  D\right)  =\left(
\begin{array}
[c]{c}%
\left(  \mathbf{u}\odot\mathbf{u}\right)  \left(  D\right)  ^{+}%
\end{array}
|%
\begin{array}
[c]{c}%
1
\end{array}
\right)  .
\end{equation}
Its relations under the shifted symplectic product are the same as
$\mathbf{u}\left(  D\right)  $:%
\begin{align}
\left(  \mathbf{a}\odot\mathbf{a}\right)  \left(  D\right)   &  =\left(
\mathbf{u}\odot\mathbf{u}\right)  \left(  D\right)  ^{-}+\left(
\mathbf{u}\odot\mathbf{u}\right)  \left(  D\right)  ^{+},\nonumber\\
&  =\left(  \mathbf{u}\odot\mathbf{u}\right)  \left(  D\right)  .
\end{align}
The vector $\mathbf{a}\left(  D\right)  $ provides a straightforward way to
make $\mathbf{N}\left(  \mathbf{u}\left(  D\right)  \right)  $ commute with
all of its shifts. We augment $\mathbf{u}\left(  D\right)  $ with
$\mathbf{a}\left(  D\right)  $. The augmented generator $\mathbf{u}^{\prime
}\left(  D\right)  $\ is as follows:%
\begin{equation}
\mathbf{u}^{\prime}\left(  D\right)  =\left(
\begin{array}
[c]{cc}%
\mathbf{z}\left(  D\right)  & \left(  \mathbf{u}\odot\mathbf{u}\right)
\left(  D\right)  ^{+}%
\end{array}
|%
\begin{array}
[c]{cc}%
\mathbf{x}\left(  D\right)  & 1
\end{array}
\right)  . \label{eq:augment-conv}%
\end{equation}
The augmented generator $\mathbf{u}^{\prime}\left(  D\right)  $ has vanishing
symplectic product because the shifted symplectic product of $\mathbf{a}%
\left(  D\right)  $ nulls the shifted symplectic product of $\mathbf{u}\left(
D\right)  $:%
\begin{equation}
\left(  \mathbf{u}^{\prime}\odot\mathbf{u}^{\prime}\right)  \left(  D\right)
=0.
\end{equation}
The augmented generator $\mathbf{N}\left(  \mathbf{u}^{\prime}\left(
D\right)  \right)  $ commutes with itself for every shift and is therefore
useful for convolutional entanglement distillation as outlined in
Section~\ref{sec:conv-ent-dist}.

We can construct an entanglement distillation protocol using an augmented
generator of this form. The first $n$ Pauli entries for every frame of
generator $\mathbf{N}\left(  \mathbf{u}^{\prime}\left(  D\right)  \right)  $
correct errors. Entry $n+1$ for every frame of $\mathbf{N}\left(
\mathbf{u}^{\prime}\left(  D\right)  \right)  $ makes $\mathbf{N}\left(
\mathbf{u}^{\prime}\left(  D\right)  \right)  $ commute with every one of its
shifts. The error-correcting properties of the code do not include errors on
the last (extra) ebit of each frame; therefore, this ebit must be noiseless.
It is necessary to catalyze the distillation procedure with $n\nu$\ noiseless
ebits where $n$ is the frame size and $\nu$ is the constraint length. The
distillation protocol requires this particular amount because it does not
correct errors and generate noiseless ebits until it has finished processing
the first basic set of generators and $\nu-1$ of its shifts. Later frames can
use the noiseless ebits generated from previous frames. Therefore these
initial noiseless ebits are negligible when calculating the yield. This
construction allows us to exploit the error-correcting properties of an
arbitrary set of Pauli matrices for a convolutional entanglement distillation
protocol.%
\begin{table}[tbp] \centering
\caption{The convolutional entanglement distillation protocol
for Example~\ref{ex:conv-ed-example-one-gen} corrects for a single-qubit error
in every fourth frame. Here we list the syndromes corresponding to
errors $X_1$, $Y_1$, and $Z_1$ on the first qubit and to errors
$X_2$, $Y_2$, and $Z_2$ on the second qubit. The syndromes are
unique so that the receiver can identify which error occurs.}\label{tbl:syndromes}%
\begin{tabular}
[c]{l|l|l|l|l|l}\hline\hline
$X_{1}$ & $Z_{1}$ & $Y_{1}$ & \thinspace$X_{2}$ & $Z_{2}$ & $Y_{2}%
$\\\hline\hline
$1$ & $0$ & $1$ & $1$ & $0$ & $1$\\
$0$ & $0$ & $0$ & $0$ & $1$ & $1$\\
$0$ & $1$ & $1$ & $1$ & $0$ & $1$\\
$1$ & $0$ & $1$ & $0$ & $0$ & $0$\\\hline\hline
\end{tabular}%
\end{table}%

We discuss the yield of such a protocol in more detail. Our construction
employs one generator with $n+1$ qubits per frame. The protocol generates $n$
noiseless ebits for every frame. But it also consumes a noiseless\ ebit for
every frame. Every frame thus produces a net of $n-1$ noiseless ebits, and the
yield of the protocol\ is $\left(  n-1\right)  /n$.

This yield of $\left(  n-1\right)  /n$\ is superior to the yield of an
entanglement distillation protocol taken from the quantum convolutional codes
of Forney et al. \cite{ieee2007forney}. Our construction should also give
entanglement distillation protocols with superior error-correcting properties
because we have no self-orthogonality constraint on the Paulis in the stabilizer.

It is possible to construct an online decoding circuit for the generator
$\mathbf{u}^{\prime}\left(  D\right)  $ by the methods given in
\cite{isit2006grassl}. A circuit satisfies the noncatastrophic property if the
polynomial entries of all of the code generators have a greatest common
divisor that is a power of the delay operator $D$ \cite{isit2006grassl}. The
online decoding circuit for this construction obeys the noncatastrophicity
property because the augmented generator $\mathbf{u}^{\prime}\left(  D\right)
$ contains 1 as one of its entries.%
\begin{figure}
[ptb]
\begin{center}
\includegraphics[
natheight=7.639800in,
natwidth=5.626500in,
height=3.7403in,
width=2.2788in
]%
{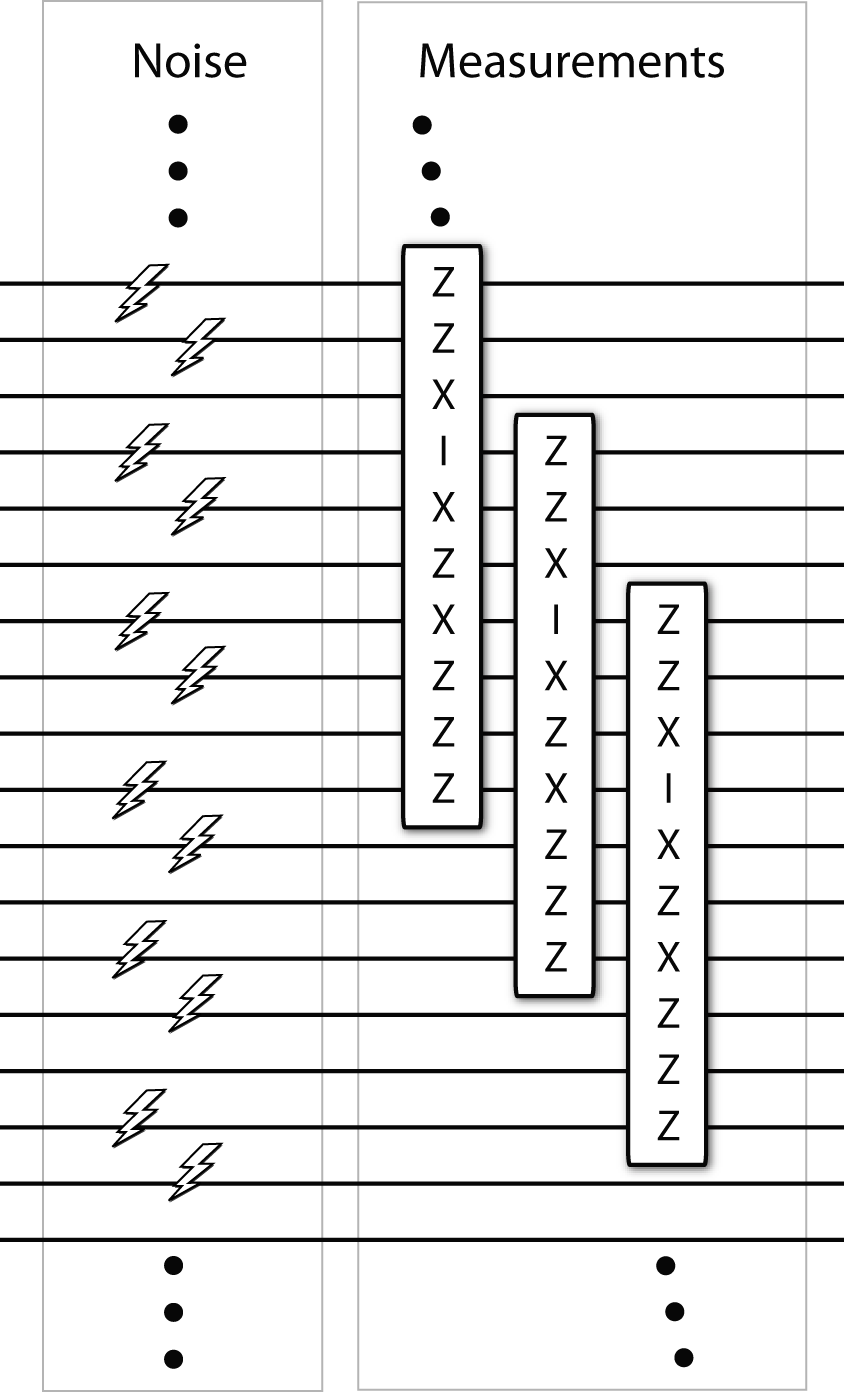}%
\caption{The above figure illustrates Bob's side of the convolutional
entanglement distillation protocol that uses entanglement assistance. The
noise affects the first and second of every three ebits that Bob shares with
Alice. Every third ebit that Alice and Bob share are noiseless. The
measurements correspond to those in Example~\ref{ex:conv-ed-example-one-gen}.}%
\label{fig:conv-dist-example}%
\end{center}
\end{figure}

\begin{example}
\label{ex:conv-ed-example-one-gen}Suppose we have the following generator%
\[
\mathbf{N}\left(  \mathbf{u}\left(  D\right)  \right)  =\left(  \cdots
|II|ZZ|IX|XZ|ZI|II|\cdots\right)  ,
\]
where%
\[
\mathbf{u}\left(  D\right)  =\left(  \left.
\begin{array}
[c]{cc}%
1+D^{3} & 1+D^{2}%
\end{array}
\right\vert
\begin{array}
[c]{cc}%
D^{2} & D
\end{array}
\right)  .
\]
The above generator corrects for an arbitrary single-qubit error in a span of
eight qubits---four frames. Table~\ref{tbl:syndromes} lists the unique
syndromes for errors in a single frame. The generator anticommutes with a
shift of itself by one or two to the left or right. The shifted symplectic
product confirms these commutation relations:%
\[
\left(  \mathbf{u\odot u}\right)  \left(  D\right)  =D+D^{2}+D^{-1}+D^{-2}.
\]
Let us follow the prescription in (\ref{eq:augment-conv}) for augmenting
generator $\mathbf{N}\left(  \mathbf{u}\left(  D\right)  \right)  $. The
following polynomial%
\begin{align}
\mathbf{a}\left(  D\right)   &  =\left(  \left.
\begin{array}
[c]{c}%
\left(  \mathbf{u\odot u}\right)  \left(  D\right)  ^{+}%
\end{array}
\right\vert
\begin{array}
[c]{c}%
1
\end{array}
\right)  ,\nonumber\\
&  =\left(  \left.
\begin{array}
[c]{c}%
D+D^{2}%
\end{array}
\right\vert
\begin{array}
[c]{c}%
1
\end{array}
\right)  ,
\end{align}
has the same commutation relations as $\mathbf{u}\left(  D\right)  $:%
\begin{equation}
\left(  \mathbf{a\odot a}\right)  \left(  D\right)  =\left(  \mathbf{u\odot
u}\right)  \left(  D\right)  .
\end{equation}
We augment $\mathbf{u}\left(  D\right)  $\ as follows:%
\[
\mathbf{u}^{\prime}\left(  D\right)  =\left(  \left.
\begin{array}
[c]{cc}%
1+D^{3} & 1+D^{2}%
\end{array}%
\begin{array}
[c]{c}%
D+D^{2}%
\end{array}
\right\vert
\begin{array}
[c]{cc}%
D^{2} & D
\end{array}%
\begin{array}
[c]{c}%
1
\end{array}
\right)  .
\]
The overall generator now looks as follows in the Pauli representation:%
\[
\mathbf{N}\left(  \mathbf{u}^{\prime}\left(  D\right)  \right)  =(\cdots
|III|ZZX|IXZ|XZZ|ZII|III|\cdots).
\]
The yield of a protocol using the above construction is 1/2.
Figure~\ref{fig:conv-dist-example} illustrates Bob's side of the protocol. It
shows which of Bob's half of the ebits are noisy and noiseless, and it gives
the measurements that Bob performs.
\end{example}

\subsection{Yield (n-m)/n Convolutional Entanglement Distillation with
Entanglement Assistance}

The construction in the above section uses only one generator for
distillation. We generalize the above construction to a code with an arbitrary
number of generators. We give an example that illustrates how to convert an
arbitrary classical quaternary convolutional code into a convolutional
entanglement distillation protocol.

Suppose we have the following $m$ generators%
\[
\left\{  \mathbf{N}\left(  \mathbf{u}_{i}\left(  D\right)  \right)  :1\leq
i\leq m\right\}  ,
\]
where%
\begin{equation}%
\begin{bmatrix}
\mathbf{u}_{1}\left(  D\right) \\
\mathbf{u}_{2}\left(  D\right) \\
\vdots\\
\mathbf{u}_{m}\left(  D\right)
\end{bmatrix}
=\left[  \left.
\begin{array}
[c]{c}%
\mathbf{z}_{1}\left(  D\right) \\
\mathbf{z}_{2}\left(  D\right) \\
\vdots\\
\mathbf{z}_{m}\left(  D\right)
\end{array}
\right\vert
\begin{array}
[c]{c}%
\mathbf{x}_{1}\left(  D\right) \\
\mathbf{x}_{2}\left(  D\right) \\
\vdots\\
\mathbf{x}_{m}\left(  D\right)
\end{array}
\right]  . \label{eq:n-k-construction-distillation}%
\end{equation}
We make no assumption about the commutation relations of the above generators.
We choose them solely for their error-correcting properties.

We again utilize the shifted symplectic product to design a convolutional
entanglement distillation protocol with multiple generators. Let us adopt the
following shorthand for the auto and cross shifted symplectic product of
generators $\mathbf{u}_{1}\left(  D\right)  ,\ldots,\mathbf{u}_{m}\left(
D\right)  $:%
\begin{align}
\mathbf{u}_{i}^{+} &  \equiv\left(  \mathbf{u}_{i}\odot\mathbf{u}_{i}\right)
\left(  D\right)  ^{+},\\
\mathbf{u}_{i,j} &  \equiv\left(  \mathbf{u}_{i}\odot\mathbf{u}_{j}\right)
\left(  D\right)  .
\end{align}
Consider the following matrix:%
\begin{equation}%
\begin{bmatrix}
\mathbf{a}_{1}\left(  D\right)  \\
\mathbf{a}_{2}\left(  D\right)  \\
\vdots\\
\mathbf{a}_{m}\left(  D\right)
\end{bmatrix}
=\left[  \left.
\begin{array}
[c]{cccc}%
\mathbf{u}_{1}^{+} & \mathbf{u}_{2,1} & \cdots & \mathbf{u}_{m,1}\\
0 & \mathbf{u}_{2}^{+} & \cdots & \mathbf{u}_{m,2}\\
\vdots &  & \ddots & \vdots\\
0 & \cdots & 0 & \mathbf{u}_{m}^{+}%
\end{array}
\right\vert \mathbf{I}_{m\times m}\right]  .\label{eq:augmented-conv-general}%
\end{equation}
The symplectic relations of the entries $\mathbf{a}_{i}\left(  D\right)  $ are
the same as the original $\mathbf{u}_{i}\left(  D\right)  $:%
\[
\left(  \mathbf{a}_{i}\odot\mathbf{a}_{j}\right)  \left(  D\right)  =\left(
\mathbf{u}_{i}\odot\mathbf{u}_{j}\right)  \left(  D\right)  \ \ \ \ \ \forall
i,j\in\left\{  1,\ldots,m\right\}  .
\]
We mention that the following matrix also has the same symplectic relations:%
\begin{equation}
\left[  \left.
\begin{array}
[c]{cccc}%
\mathbf{u}_{1}^{+} & 0 & \cdots & 0\\
\mathbf{u}_{1,2} & \mathbf{u}_{2}^{+} & \cdots & \vdots\\
\vdots &  & \ddots & 0\\
\mathbf{u}_{1,m} & \mathbf{u}_{2,m} & \cdots & \mathbf{u}_{m}^{+}%
\end{array}
\right\vert \mathbf{I}_{m\times m}\right]  .\label{eq:alternate-constr-matrix}%
\end{equation}
Let us rewrite (\ref{eq:augmented-conv-general}) as follows:%
\begin{equation}%
\begin{bmatrix}
\mathbf{a}_{1}\left(  D\right)  \\
\mathbf{a}_{2}\left(  D\right)  \\
\vdots\\
\mathbf{a}_{m}\left(  D\right)
\end{bmatrix}
=\left[  \left.
\begin{array}
[c]{c}%
\mathbf{z}_{1}^{\prime}\left(  D\right)  \\
\mathbf{z}_{2}^{\prime}\left(  D\right)  \\
\vdots\\
\mathbf{z}_{m}^{\prime}\left(  D\right)
\end{array}
\right\vert
\begin{array}
[c]{c}%
\mathbf{x}_{1}^{\prime}\left(  D\right)  \\
\mathbf{x}_{2}^{\prime}\left(  D\right)  \\
\vdots\\
\mathbf{x}_{m}^{\prime}\left(  D\right)
\end{array}
\right]  .
\end{equation}
The above matrix provides a straightforward way to make the original
generators commute with all of their shifts. We augment the generators in
(\ref{eq:n-k-construction-distillation}) by the generators $\mathbf{a}%
_{i}\left(  D\right)  $\ to get the following $m\times2\left(  n+m\right)
$\ matrix:%
\begin{align}
&  \mathbf{U}^{\prime}\left(  D\right)  =\left[  \left.
\begin{array}
[c]{c}%
\mathbf{Z}\left(  D\right)
\end{array}
\right\vert
\begin{array}
[c]{c}%
\mathbf{X}\left(  D\right)
\end{array}
\right]  =\nonumber\\
&  \left[  \left.
\begin{array}
[c]{cc}%
\mathbf{z}_{1}\left(  D\right)   & \mathbf{z}_{1}^{\prime}\left(  D\right)  \\
\mathbf{z}_{2}\left(  D\right)   & \mathbf{z}_{2}^{\prime}\left(  D\right)  \\
\vdots & \vdots\\
\mathbf{z}_{m}\left(  D\right)   & \mathbf{z}_{m}^{\prime}\left(  D\right)
\end{array}
\right\vert
\begin{array}
[c]{cc}%
\mathbf{x}_{1}\left(  D\right)   & \mathbf{x}_{1}^{\prime}\left(  D\right)  \\
\mathbf{x}_{2}\left(  D\right)   & \mathbf{x}_{2}^{\prime}\left(  D\right)  \\
\vdots & \vdots\\
\mathbf{x}_{m}\left(  D\right)   & \mathbf{x}_{m}^{\prime}\left(  D\right)
\end{array}
\right]  .
\end{align}
Every row of the augmented matrix $\mathbf{U}^{\prime}\left(  D\right)  $ has
vanishing symplectic product with itself and any other row. This condition is
equivalent to the following matrix condition for shifted symplectic
orthogonality \cite{arxiv2004olliv}:%
\begin{equation}
\mathbf{Z}\left(  D^{-1}\right)  \mathbf{X}\left(  D\right)  ^{T}%
-\mathbf{X}\left(  D^{-1}\right)  \mathbf{Z}\left(  D\right)  ^{T}=0.
\end{equation}
The construction gives a commuting set of generators for arbitrary shifts and
thus forms a valid stabilizer.

We can readily develop a convolutional entanglement distillation protocol
using the above formulation. The generators in the augmented matrix
$\mathbf{U}^{\prime}\left(  D\right)  $\ correct for errors on the first $n$
ebits. The last $m$ ebits are noiseless ebits that help to obtain a commuting
stabilizer. It is necessary to catalyze the distillation protocol with
$\left(  n+m\right)  \nu$ noiseless ebits. Later frames can use the noiseless
ebits generated from previous frames. These initial noiseless ebits are
negligible when calculating the yield.

We comment more on the yield of the protocol. The protocol requires a set of
$m$ generators with $n+m$ Pauli entries. It generates $n$ ebits for every
frame. But it consumes $m$ noiseless ebits per frame. The net yield of a
protocol using the above construction is thus $\left(  n-m\right)  /n$.

The key benefit of the above construction is that we can use an arbitrary set
of Paulis for distilling noiseless ebits. This arbitrariness in the Paulis
implies that we can import an arbitrary classical convolutional binary or
quaternary code for use in a convolutional entanglement distillation protocol.

It is again straightforward to develop a noncatastrophic decoding circuit
using previous techniques \cite{isit2006grassl}. Every augmented generator in
$\mathbf{U}^{\prime}\left(  D\right)  $ has 1 as an entry so that it satisfies
the property required for noncatastrophicity.

\begin{example}
We begin with a classical quaternary convolutional code with entries from
$\mathbb{F}_{4}$:%
\begin{equation}
\left(  \cdots|0000|1\bar{\omega}10|1101|0000|\cdots\right)  .
\end{equation}
The above code is a convolutional version of the classical quaternary block
code from Ref.~\cite{science2006brun}. We multiply the above generator by
$\bar{\omega}$ and $\omega$ as prescribed in Refs.
\cite{ieee1998calderbank,ieee2007forney} and use the following map,%
\begin{equation}%
\begin{tabular}
[c]{|l|l|}\hline
$\mathbb{F}_{4}$ & $\Pi$\\\hline
$0$ & $I$\\
$\omega$ & $X$\\
$1$ & $Y$\\
$\bar{\omega}$ & $Z$\\\hline
\end{tabular}
\ \ \ \ ,
\end{equation}
to obtain the following Pauli generators%
\begin{align}
\mathbf{N}\left(  \mathbf{u}_{1}\left(  D\right)  \right)   &  =\left(
\cdots|IIII|ZXZI|ZZIZ|IIII|\cdots\right)  ,\nonumber\\
\mathbf{N}\left(  \mathbf{u}_{2}\left(  D\right)  \right)   &  =\left(
\cdots|IIII|XYXI|XXIX|IIII|\cdots\right)  .
\end{align}
We determine binary polynomials corresponding to the above Pauli generators:%
\begin{multline}
\left(
\begin{array}
[c]{c}%
\mathbf{u}_{1}\left(  D\right) \\
\mathbf{u}_{2}\left(  D\right)
\end{array}
\right)  =\\
\left(  \left.
\begin{array}
[c]{cccc}%
1+D & D & 1 & D\\
0 & 1 & 0 & 0
\end{array}
\right\vert
\begin{array}
[c]{cccc}%
0 & 1 & 0 & 0\\
1+D & 1+D & 1 & D
\end{array}
\right)  .
\end{multline}
The first generator anticommutes with itself shifted by one to the left or
right, the second generator anticommutes with itself shifted by one to the
left or right, and the first generator anticommutes with the second shifted by
one to the left. The following shifted symplectic products confirm the above
commutation relations:%
\begin{align}
\left(  \mathbf{u}_{1}\odot\mathbf{u}_{1}\right)  \left(  D\right)   &
=D^{-1}+D,\nonumber\\
\left(  \mathbf{u}_{2}\odot\mathbf{u}_{2}\right)  \left(  D\right)   &
=D^{-1}+D,\nonumber\\
\left(  \mathbf{u}_{1}\odot\mathbf{u}_{2}\right)  \left(  D\right)   &  =D.
\label{eq:symp-prod-relations}%
\end{align}
Consider the following two generators:%
\begin{equation}
\left(
\begin{array}
[c]{c}%
\mathbf{a}_{1}\left(  D\right) \\
\mathbf{a}_{2}\left(  D\right)
\end{array}
\right)  =\left(  \left.
\begin{array}
[c]{cc}%
D & 0\\
D & D
\end{array}
\right\vert
\begin{array}
[c]{cc}%
1 & 0\\
0 & 1
\end{array}
\right)  .
\end{equation}
Their relations under the shifted symplectic product are the same as those in
(\ref{eq:symp-prod-relations}).%
\begin{align}
\left(  \mathbf{a}_{1}\odot\mathbf{a}_{1}\right)  \left(  D\right)   &
=\left(  \mathbf{u}_{1}\odot\mathbf{u}_{1}\right)  \left(  D\right)
,\nonumber\\
\left(  \mathbf{a}_{2}\odot\mathbf{a}_{2}\right)  \left(  D\right)   &
=\left(  \mathbf{u}_{2}\odot\mathbf{u}_{2}\right)  \left(  D\right)
,\nonumber\\
\left(  \mathbf{a}_{1}\odot\mathbf{a}_{2}\right)  \left(  D\right)   &
=\left(  \mathbf{u}_{1}\odot\mathbf{u}_{2}\right)  \left(  D\right)  .
\end{align}
We use the construction from (\ref{eq:alternate-constr-matrix}) so that we
have positive delay operators in the augmented matrix. We augment the
generators $\mathbf{u}_{1}\left(  D\right)  $\ and $\mathbf{u}_{2}\left(
D\right)  $\ to generators $\mathbf{u}_{1}^{\prime}\left(  D\right)  $\ and
$\mathbf{u}_{2}^{\prime}\left(  D\right)  $\ respectively as follows. The
augmented \textquotedblleft Z matrix\textquotedblright\ is%
\begin{equation}
\mathbf{Z}\left(  D\right)  =\left(
\begin{array}
[c]{cccc}%
1+D & D & 1 & D\\
0 & 1 & 0 & 0
\end{array}%
\begin{array}
[c]{cc}%
D & 0\\
D & D
\end{array}
\right)  ,
\end{equation}
and the augmented \textquotedblleft X matrix\textquotedblright\ is%
\begin{equation}
\mathbf{X}\left(  D\right)  =\left(
\begin{array}
[c]{cccc}%
0 & 1 & 0 & 0\\
1+D & 1+D & 1 & D
\end{array}%
\begin{array}
[c]{cc}%
1 & 0\\
0 & 1
\end{array}
\right)  .
\end{equation}
The augmented matrix $\mathbf{U}^{\prime}\left(  D\right)  $ is%
\begin{equation}
\mathbf{U}^{\prime}\left(  D\right)  =\left[  \left.
\begin{array}
[c]{c}%
\mathbf{Z}\left(  D\right)
\end{array}
\right\vert
\begin{array}
[c]{c}%
\mathbf{X}\left(  D\right)
\end{array}
\right]  .
\end{equation}
The first row of $\mathbf{U}^{\prime}\left(  D\right)  $ is generator
$\mathbf{u}_{1}^{\prime}\left(  D\right)  $\ and the second row is
$\mathbf{u}_{2}^{\prime}\left(  D\right)  $. The augmented generators have the
following Pauli representation.%
\begin{multline}
\mathbf{N}\left(  \mathbf{u}_{1}^{\prime}\left(  D\right)  \right)
=\nonumber\\
\left(  \cdots|IIIIII|ZXZIXI|ZZIZZI|IIIIII|\cdots\right)  ,
\end{multline}%
\begin{multline}
\mathbf{N}\left(  \mathbf{u}_{2}^{\prime}\left(  D\right)  \right)
=\nonumber\\
\left(  \cdots|IIIIII|XYXIIX|XXIXZZ|IIIIII|\cdots\right)  .
\end{multline}
The original block code from Ref.~\cite{science2006brun} corrects for an
arbitrary single-qubit error. The above entanglement distillation protocol
corrects for a single-qubit error in eight qubits---two frames. This
error-correcting capability follows from the capability of the block code. The
yield of a protocol using the above stabilizer is again 1/2.
\end{example}

\subsection{CSS-Like Construction for a Convolutional Entanglement
Distillation Protocol}

We finally present a construction that allows us to import two arbitrary
binary classical codes for use in a convolutional entanglement distillation
protocol. The construction is similar to a CSS\ code because one code corrects
for bit flips and the other corrects for phase flips.

We could simply use the technique from the previous section to construct a
convolutional entanglement-distillation protocol. We could represent both
classical codes as codes over $\mathbb{F}_{4}$. We could multiply the bit-flip
code by $\omega$ and the phase-flip code by $\bar{\omega}$ and use the above
map from $\mathbb{F}_{4}$ to the Paulis. We could then use the above method
for augmentation and obtain a valid quantum code for entanglement
distillation. But there is a better method that exploits the structure of a
CSS\ code to minimize the number of initial catalytic noiseless ebits.

Our algorithm below uses a Gram-Schmidt like orthogonalization procedure to
minimize the number of initial noiseless ebits. The procedure is similar to
the algorithm in \cite{arx2006brun}\ with some key differences.

Suppose we have $m$ generators $\left\{  \mathbf{N}\left(  \mathbf{w}%
_{i}\left(  D\right)  \right)  :1\leq i\leq m\right\}  $ where%
\begin{equation}%
\begin{bmatrix}
\mathbf{w}_{1}\left(  D\right) \\
\vdots\\
\mathbf{w}_{p}\left(  D\right) \\
\mathbf{w}_{p+1}\left(  D\right) \\
\vdots\\
\mathbf{w}_{m}\left(  D\right)
\end{bmatrix}
=\left[  \left.
\begin{array}
[c]{c}%
\mathbf{z}_{1}\left(  D\right) \\
\vdots\\
\mathbf{z}_{p}\left(  D\right) \\
\mathbf{0}\\
\vdots\\
\mathbf{0}%
\end{array}
\right\vert
\begin{array}
[c]{c}%
\mathbf{0}\\
\vdots\\
\mathbf{0}\\
\mathbf{x}_{1}\left(  D\right) \\
\vdots\\
\mathbf{x}_{m-p}\left(  D\right)
\end{array}
\right]  .
\end{equation}
and each vector $\mathbf{w}_{i}\left(  D\right)  $ has length $2n$. The above
matrix could come from two binary classical codes. The vectors $\mathbf{z}%
_{1}\left(  D\right)  $,\ldots,$\mathbf{z}_{p}\left(  D\right)  $ could come
from one code, and the vectors $\mathbf{x}_{1}\left(  D\right)  $%
,\ldots,$\mathbf{x}_{m-p}\left(  D\right)  $ could come from another code. The
following orthogonality relations hold for the above vectors:%
\begin{align}
\forall\ \ 1\leq i,j\leq p  &  :\left(  \mathbf{w}_{i}\odot\mathbf{w}%
_{j}\right)  \left(  D\right)  =0,\\
\forall\ \ p+1\leq i^{\prime},j^{\prime}\leq m  &  :\left(  \mathbf{w}%
_{i^{\prime}}\odot\mathbf{w}_{j^{\prime}}\right)  \left(  D\right)  =0.
\end{align}
We exploit the above orthogonality relations in the algorithm below.

We can perform a Gram-Schmidt process on the above set of vectors. This
process orthogonalizes the vectors with respect to the shifted symplectic
product. The procedure does not change the error-correcting properties of the
original codes because all operations are linear.

The algorithm breaks the set of vectors above into pairs. Each pair consists
of two vectors which are symplectically nonorthogonal to each other, but which
are symplectically orthogonal to all other pairs. Any remaining vectors that
are symplectically orthogonal to all other vectors are collected into a
separate set, which we call the set of isotropic vectors. This idea is similar
to the decomposition of a vector space into an isotropic and symplectic part.
We cannot label the decomposition as such because the shifted symplectic
product is not a true symplectic product.

We detail the initialization of the algorithm. Set parameters $i=0$, $c=0$,
$l=0$. The index $i$ labels the total number of vectors processed, $c$ gives
the number of pairs, and $l$ labels the number of vectors with no partner.
Initialize sets $\mathcal{U}$ and $\mathcal{V}$ to be null: $\mathcal{U}%
=\mathcal{V}=\emptyset$. $\mathcal{U}$ keeps track of the pairs and
$\mathcal{V}$ keeps track of the vectors with no partner.

The algorithm proceeds as follows. While $i\leq m$, let $j\geq2c+l+2$ be the
smallest index for a $\mathbf{w}_{j}\left(  D\right)  $ for which $\left(
\mathbf{w}_{2c+l+1}\odot\mathbf{w}_{j}\right)  \left(  D\right)  \neq0$.
Increment $l$ and $i$ by one, add $i$ to $\mathcal{V}$, and proceed to the
next round if no such pair exists. Otherwise, swap $\mathbf{w}_{j}\left(
D\right)  $ with $\mathbf{w}_{2c+l+2}\left(  D\right)  $. For $r\in\left\{
2c+l+3,\ldots,m\right\}  $, perform%
\begin{multline}
\mathbf{w}_{r}\left(  D\right)  =\left(  \mathbf{w}_{2c+l+2}\odot
\mathbf{w}_{2c+l+1}\right)  \left(  D\right)  \mathbf{w}_{r}\left(  D\right)
\nonumber\\
+\left(  \mathbf{w}_{r}\odot\mathbf{w}_{2c+l+2}\right)  \left(  D^{-1}\right)
\mathbf{w}_{2c+l+1}\left(  D\right)  .
\end{multline}
if $\mathbf{w}_{r}\left(  D\right)  $ has a purely $z$ component. Perform%
\begin{multline}
\mathbf{w}_{r}\left(  D\right)  =\left(  \mathbf{w}_{2c+l+1}\odot
\mathbf{w}_{2c+l+2}\right)  \left(  D\right)  \mathbf{w}_{r}\left(  D\right)
\nonumber\\
+\left(  \mathbf{w}_{r}\odot\mathbf{w}_{2c+l+1}\right)  \left(  D^{-1}\right)
\mathbf{w}_{2c+l+2}\left(  D\right)  .
\end{multline}
if $\mathbf{w}_{r}\left(  D\right)  $ has a purely $x$ component. Divide every
element in $\mathbf{w}_{r}\left(  D\right)  $ by the greatest common factor if
the GCF is not equal to one. Then%
\begin{equation}
\left(  \mathbf{w}_{r}\odot\mathbf{w}_{2c+l+1}\right)  \left(  D\right)
=\left(  \mathbf{w}_{r}\odot\mathbf{w}_{2c+l+2}\right)  \left(  D\right)  =0.
\end{equation}
Increment $c$ by one, increment $i$ by one, add $i$ to $\mathcal{U}$, and
increment $i$ by one. Proceed to the next round.

We now give the method for augmenting the above generators so that they form a
commuting stabilizer. At the end of the algorithm, the sets $\mathcal{U}$ and
$\mathcal{V}$\ have the following sizes: $\left\vert \mathcal{U}\right\vert
=c$ and $\left\vert \mathcal{V}\right\vert =l$. Let us relabel the vectors
$\mathbf{w}_{i}\left(  D\right)  $ for all $1\leq i\leq2c+l$. We relabel all
pairs: call the first $\mathbf{u}_{i}\left(  D\right)  $ and call its partner
$\mathbf{v}_{i}\left(  D\right)  $ for all $1\leq i\leq c$. Call any vector
without a partner $\mathbf{u}_{c+i}\left(  D\right)  $ for all $1\leq i\leq
l$. The relabeled vectors have the following shifted symplectic product
relations after the Gram-Schmidt procedure:%
\begin{align}
\left(  \mathbf{u}_{i}\odot\mathbf{v}_{j}\right)  \left(  D\right)   &
=f_{i}\left(  D\right)  \delta_{ij}\ \ \forall\ \ i,j\in\left\{
1,\ldots,c\right\}  ,\nonumber\\
\left(  \mathbf{u}_{i}\odot\mathbf{u}_{j}\right)  \left(  D\right)   &
=0\ \ \ \ \ \ \ \ \ \ \ \ \forall\ \ i,j\in\left\{  1,\ldots,l\right\}
,\nonumber\\
\left(  \mathbf{v}_{i}\odot\mathbf{v}_{j}\right)  \left(  D\right)   &
=0\ \ \ \ \ \ \ \ \ \ \ \ \forall\ \ i,j\in\left\{  1,\ldots,c\right\}  ,
\end{align}
where $f_{i}\left(  D\right)  $ is an arbitrary polynomial. Let us arrange the
above generators in a matrix as follows:%
\begin{equation}%
\begin{bmatrix}
\mathbf{u}_{1}\left(  D\right) \\
\vdots\\
\mathbf{u}_{c}\left(  D\right) \\
\mathbf{v}_{1}\left(  D\right) \\
\vdots\\
\mathbf{v}_{c}\left(  D\right) \\
\mathbf{u}_{c+1}\left(  D\right) \\
\vdots\\
\mathbf{u}_{c+l}\left(  D\right)
\end{bmatrix}
.
\end{equation}
We augment the above generators with the following matrix so that all vectors
are orthogonal to each other:%
\begin{equation}
\left[  \left.
\begin{array}
[c]{cccc}%
f_{1}\left(  D^{-1}\right)  & 0 & \cdots & 0\\
0 & f_{2}\left(  D^{-1}\right)  &  & \vdots\\
\vdots &  & \ddots & 0\\
0 & \cdots & 0 & f_{c}\left(  D^{-1}\right) \\
\mathbf{0}_{c\times1} & \mathbf{0}_{c\times1} & \cdots & \mathbf{0}_{c\times
1}\\
\mathbf{0}_{l\times1} & \mathbf{0}_{l\times1} & \cdots & \mathbf{0}_{l\times1}%
\end{array}
\right\vert
\begin{array}
[c]{c}%
\mathbf{0}_{1\times c}\\
\mathbf{0}_{1\times c}\\
\vdots\\
\mathbf{0}_{1\times c}\\
\mathbf{I}_{c\times c}\\
\mathbf{0}_{l\times c}%
\end{array}
\right]  .
\end{equation}

The yield of a protocol using the above construction is $\left(  n-m\right)
/n$. Suppose we use an $\left[  n,k_{1}\right]  $ classical binary
convolutional code for the bit flips and an $\left[  n,k_{2}\right]  $
classical binary convolutional code for the phase flips. Then the
convolutional entanglement distillation protocol has yield $\left(
k_{1}+k_{2}-n\right)  /n$.

\begin{example}
Consider a binary classical convolutional code with the following parity check
matrix:%
\begin{equation}%
\begin{bmatrix}
1+D & D & 1
\end{bmatrix}
.
\end{equation}
We can use the above parity check matrix to correct both bit and phase flip
errors in an entanglement distillation protocol. Our initial quantum parity
check matrix is%
\begin{equation}
\left[  \left.
\begin{array}
[c]{ccc}%
1+D & D & 1\\
0 & 0 & 0
\end{array}
\right\vert
\begin{array}
[c]{ccc}%
0 & 0 & 0\\
1+D & D & 1
\end{array}
\right]  .
\end{equation}
The shifted symplectic product for the first and second row is $D^{-1}+D$. We
therefore augment the above matrix as follows:%
\begin{equation}
\left[  \left.
\begin{array}
[c]{cccc}%
1+D & D & 1 & D^{-1}+D\\
0 & 0 & 0 & 0
\end{array}
\right\vert
\begin{array}
[c]{cccc}%
0 & 0 & 0 & 0\\
1+D & D & 1 & 1
\end{array}
\right]  .
\end{equation}
The above matrix gives a valid stabilizer for use in an entanglement
distillation protocol. The yield of a protocol using the above stabilizer is 1/3.
\end{example}

\section{Conclusion and Current Work}

We constructed a theory of convolutional entanglement distillation. The
entanglement-assisted protocol assumes that the sender and receiver have some
noiseless ebits to use as a catalyst for distilling more ebits. These
protocols have the benefit of lifting the self-orthogonality constraint. Thus
we are able to import an arbitrary classical convolutional code for use in a
convolutional entanglement distillation protocol. The error-correcting
properties and rate of the classical code translate to the quantum case. Brun,
Devetak, and Hsieh first constructed the method for importing an arbitrary
classical block code in their work on entanglement-assisted codes
\cite{arx2006brun,science2006brun}. Our theory of convolutional entanglement
distillation paves the way for exploring protocols that approach the optimal
distillable entanglement by using the well-established theory of classical
convolutional coding.

Convolutional entanglement distillation protocols also hold some key
advantages over block entanglement distillation protocols. They have a higher
yield of ebits, lower decoding complexity, and are an online protocol that a
sender and receiver can employ as they acquire more noisy ebits.

We suggest that convolutional entanglement distillation protocols may bear
some advantages for distillation of a secret key because of the strong
connection between distillation and privacy \cite{PhysRevLett.85.441}. We are
currently investigating whether convolutional entanglement distillation
protocols can improve the secret key rate for quantum key distribution.

The authors thank Igor Devetak and Zhicheng Luo for useful discussions and
thank Saikat Guha for locating a copy of Jonsson's master's thesis.
MMW\ acknowledges support from NSF Grant 0545845,\ and HK and TAB\ acknowledge
support from NSF Grant CCF-0448658. All authors are grateful to the Hearne
Insitute for Theoretical Physics for hosting MMW\ as a visiting researcher.

\section{Appendix}

\begin{example}
\label{ex:science-code}We present an example of an entanglement-assisted code
that corrects an arbitrary single-qubit error \cite{science2006brun}. Suppose
the sender wants to use the quantum error-correcting properties of the
following nonabelian subgroup of $\Pi^{4}$:%
\begin{equation}%
\begin{array}
[c]{cccc}%
Z & X & Z & I\\
Z & Z & I & Z\\
X & Y & X & I\\
X & X & I & X
\end{array}
\label{eq:science-example}%
\end{equation}
The first two generators anticommute. We obtain a modified third generator by
multiplying the third generator by the second. We then multiply the last
generator by the first, second, and modified third generators. The
error-correcting properties of the generators are invariant under these
operations. The modified generators are as follows:%
\begin{equation}%
\begin{array}
[c]{cccccc}%
g_{1} & = & Z & X & Z & I\\
g_{2} & = & Z & Z & I & Z\\
g_{3} & = & Y & X & X & Z\\
g_{4} & = & Z & Y & Y & X
\end{array}
\end{equation}
The above set of generators have the commutation relations given by the
fundamental theorem of symplectic geometry:%
\begin{align*}
&  \left\{  g_{1},g_{2}\right\}  =\left[  g_{1},g_{3}\right]  =\left[
g_{1},g_{4}\right]  ,\\
&  =\left[  g_{2},g_{3}\right]  =\left[  g_{2},g_{4}\right]  =\left[
g_{3},g_{4}\right]  =0.
\end{align*}
The above set of generators is unitarily equivalent to the following canonical
generators:%
\begin{equation}%
\begin{array}
[c]{cccc}%
X & I & I & I\\
Z & I & I & I\\
I & Z & I & I\\
I & I & Z & I
\end{array}
\label{eq:canonical-Paulis}%
\end{equation}
We can add one ebit to resolve the anticommutativity of the first two
generators:%
\begin{equation}
\left.
\begin{array}
[c]{cccc}%
X & I & I & I\\
Z & I & I & I\\
I & Z & I & I\\
I & I & Z & I
\end{array}
\right\vert
\begin{array}
[c]{c}%
X\\
Z\\
I\\
I
\end{array}
\label{eq:canonical-stabilizer}%
\end{equation}
The following state is an eigenstate of the above stabilizer%
\begin{equation}
\left\vert \Phi^{+}\right\rangle ^{AB}\left\vert 00\right\rangle
^{A}\left\vert \psi\right\rangle ^{A}. \label{eq:canonical-state}%
\end{equation}
where $\left\vert \psi\right\rangle ^{A}$ is a qubit that the sender wants to
encode. The encoding unitary then rotates the generators in
(\ref{eq:canonical-stabilizer})\ to the following set of globally commuting
generators:%
\begin{equation}
\left.
\begin{array}
[c]{cccc}%
Z & X & Z & I\\
Z & Z & I & Z\\
Y & X & X & Z\\
Z & Y & Y & X
\end{array}
\right\vert
\begin{array}
[c]{c}%
X\\
Z\\
I\\
I
\end{array}
\label{eq:encoded-stabilizer}%
\end{equation}
The receiver measures the above generators upon receipt of all qubits to
detect and correct errors.
\end{example}

\subsection{Encoding Algorithm}

\label{sec:encoding-alg-ent-assist}We continue with the previous example. We
detail an algorithm for determining an encoding circuit and the optimal number
of ebits for the entanglement-assisted code. The operators in
(\ref{eq:science-example}) have the following representation as a binary
matrix:%
\begin{equation}
H=\left[  \left.
\begin{array}
[c]{cccc}%
1 & 0 & 1 & 0\\
1 & 1 & 0 & 1\\
0 & 1 & 0 & 0\\
0 & 0 & 0 & 0
\end{array}
\right\vert
\begin{array}
[c]{cccc}%
0 & 1 & 0 & 0\\
0 & 0 & 0 & 0\\
1 & 1 & 1 & 0\\
1 & 1 & 0 & 1
\end{array}
\right]  .
\end{equation}
Call the matrix to the left of the vertical bar the \textquotedblleft$Z$
matrix\textquotedblright\ and the matrix to the right of the vertical bar the
\textquotedblleft$X$ matrix.\textquotedblright

The algorithm consists of row and column operations on the above matrix. Row
operations do not affect the error-correcting properties of the code but are
crucial for arriving at the optimal decomposition from the fundamental theorem
of symplectic geometry. The operations available for manipulating columns of
the above matrix are Clifford operations \cite{thesis97gottesman}. Clifford
operations preserve the Pauli group $\Pi^{n}$ under conjugation.\ The
CNOT\ gate, the Hadamard gate, and the Phase gate generate the Clifford group.
A CNOT\ gate from qubit $i$ to qubit $j$ adds column $i$ to column $j$ in the
$X$ matrix and adds column $j$ to column $i$ in the $Z$ matrix. A Hadamard
gate on qubit $i$ swaps column $i$ in the $Z$ matrix with column $i$ in the
$X$ matrix and vice versa. A phase gate on qubit $i$\ adds column $i$ in the
$X$ matrix to column $i$ in the $Z$ matrix. Three CNOT\ gates implement a
qubit swap operation \cite{book2000mikeandike}. The effect of a swap on qubits
$i$ and $j$ is to swap columns $i$ and $j$ in both the $X$ and $Z$ matrix.

The algorithm begins by computing the symplectic product between the first row
and all other rows. We emphasize that the symplectic product here is the
standard symplectic product. Leave the matrix as it is if the first row is not
symplectically orthogonal to the second row or if the first row is
symplectically orthogonal to all other rows. Otherwise, swap the second row
with the first available row that is not symplectically orthogonal to the
first row. In our example, the first row is not symplectically orthogonal to
the second so we leave all rows as they are.

Arrange the first row so that the top left entry in the $X$ matrix is one. A
CNOT, swap, Hadamard, or combinations of these operations can achieve this
result. We can have this result in our example by swapping qubits one and two.
The matrix becomes%
\begin{equation}
\left[  \left.
\begin{array}
[c]{cccc}%
0 & 1 & 1 & 0\\
1 & 1 & 0 & 1\\
1 & 0 & 0 & 0\\
0 & 0 & 0 & 0
\end{array}
\right\vert
\begin{array}
[c]{cccc}%
1 & 0 & 0 & 0\\
0 & 0 & 0 & 0\\
1 & 1 & 1 & 0\\
1 & 1 & 0 & 1
\end{array}
\right]  .
\end{equation}

Perform CNOTs to clear the entries in the $X$ matrix in the top row to the
right of the leftmost entry. These entries are already zero in this example so
we need not do anything. Proceed to the clear the entries in the first row of
the $Z$ matrix. Perform a phase gate to clear the leftmost entry in the first
row of the $Z$ matrix if it is equal to one. It is equal to zero in this case
so we need not do anything. We then use Hadamards and CNOTs to clear the other
entries in the first row of the $Z$ matrix.

We perform the above operations for our example. Perform a Hadamard on qubits
two and three. The matrix becomes%
\begin{equation}
\left[  \left.
\begin{array}
[c]{cccc}%
0 & 0 & 0 & 0\\
1 & 0 & 0 & 1\\
1 & 1 & 1 & 0\\
0 & 1 & 0 & 0
\end{array}
\right\vert
\begin{array}
[c]{cccc}%
1 & 1 & 1 & 0\\
0 & 1 & 0 & 0\\
1 & 0 & 0 & 0\\
1 & 0 & 0 & 1
\end{array}
\right]  .
\end{equation}
Perform a CNOT\ from qubit one to qubit two and from qubit one to qubit three.
The matrix becomes%
\begin{equation}
\left[  \left.
\begin{array}
[c]{cccc}%
0 & 0 & 0 & 0\\
1 & 0 & 0 & 1\\
1 & 1 & 1 & 0\\
1 & 1 & 0 & 0
\end{array}
\right\vert
\begin{array}
[c]{cccc}%
1 & 0 & 0 & 0\\
0 & 1 & 0 & 0\\
1 & 1 & 1 & 0\\
1 & 1 & 1 & 1
\end{array}
\right]  .
\end{equation}
The first row is complete. We now proceed to clear the entries in the second
row. Perform a Hadamard on qubits one and four. The matrix becomes%
\begin{equation}
\left[  \left.
\begin{array}
[c]{cccc}%
1 & 0 & 0 & 0\\
0 & 0 & 0 & 0\\
1 & 1 & 1 & 0\\
1 & 1 & 0 & 1
\end{array}
\right\vert
\begin{array}
[c]{cccc}%
0 & 0 & 0 & 0\\
1 & 1 & 0 & 1\\
1 & 1 & 1 & 0\\
1 & 1 & 1 & 0
\end{array}
\right]  .
\end{equation}
Perform a CNOT\ from qubit one to qubit two and from qubit one to qubit four.
The matrix becomes%
\begin{equation}
\left[  \left.
\begin{array}
[c]{cccc}%
1 & 0 & 0 & 0\\
0 & 0 & 0 & 0\\
0 & 1 & 1 & 0\\
1 & 1 & 0 & 1
\end{array}
\right\vert
\begin{array}
[c]{cccc}%
0 & 0 & 0 & 0\\
1 & 0 & 0 & 0\\
1 & 0 & 1 & 1\\
1 & 0 & 1 & 1
\end{array}
\right]  .
\end{equation}
The first two rows are now complete. They need one ebit to compensate for
their anticommutativity or their nonorthogonality with respect to the
symplectic product.

Now we perform a \textquotedblleft Gram-Schmidt
orthogonalization\textquotedblright\ with respect to the symplectic product.
Add row 1 to any other row that has one as the leftmost entry in its $Z$
matrix. Add row two to any other row that has one as the leftmost entry in its
$X$ matrix. For our example, we add row one to row four and we add row two to
rows three and four. The matrix becomes%
\begin{equation}
\left[  \left.
\begin{array}
[c]{cccc}%
1 & 0 & 0 & 0\\
0 & 0 & 0 & 0\\
0 & 1 & 1 & 0\\
0 & 1 & 0 & 1
\end{array}
\right\vert
\begin{array}
[c]{cccc}%
0 & 0 & 0 & 0\\
1 & 0 & 0 & 0\\
0 & 0 & 1 & 1\\
0 & 0 & 1 & 1
\end{array}
\right]  .
\end{equation}
The first two rows are now symplectically orthogonal to all other rows per the
fundamental theorem of symplectic geometry.%

\begin{figure}
[ptb]
\begin{center}
\[
\Qcircuit@C=0.5em @R=1.0em  {
&  &  & \qw& \qw& \qw& \qw& \qw& \qw& \qw& \qw& \qw& \qw& \qw\gategroup{1}%
{3}{2}{3}{1.0em}{\{}\\
& \lstick{\raisebox{2em}{$\ket{\Phi^{+}}^{BA}$}} &  & \qw& \qw& \qw& \qw
& \qw& \ctrl{1} \qwx[3] & \gate{H} & \ctrl{1} \qwx[2] & \qw& \qswap
\qwx[1] & \qw\\
& \lstick{\ket{0}^A}      & \gate{H} & \qw& \ctrl{1} & \gate{P} & \ctrl{1}
\qwx[2] & \gate{H} & \targ& \qw& \targ& \gate{H} & \qswap& \qw\\
& \lstick{\ket{0}^A}      & \gate{H} & \ctrl{1} & \targ& \gate{H} & \targ
& \qw& \qw& \qw& \targ& \gate{H} & \qw& \qw\\
& \lstick{\ket{\psi}^A}  & \qw& \targ& \gate{H} & \qw& \targ& \qw
& \targ& \gate{H} & \qw& \qw& \qw& \qw}
\]
\end{center}
\caption{Encoding circuit for the entanglement-assisted code from \cite
{science2006brun}%
. The ``H'' gate is a Hadamard gate and the ``P'' gate is a phase gate.}
\label{fig:encoding-circuit-science}
\end{figure}
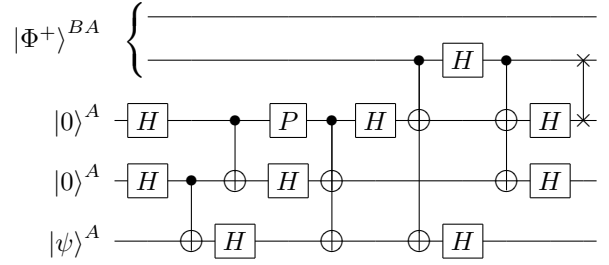%
We proceed with the same algorithm on the next two rows. The next two rows are
symplectically orthogonal to each other so we can deal with them individually.
Perform a Hadamard on qubit two. The matrix becomes%
\begin{equation}
\left[  \left.
\begin{array}
[c]{cccc}%
1 & 0 & 0 & 0\\
0 & 0 & 0 & 0\\
0 & 0 & 1 & 0\\
0 & 0 & 0 & 1
\end{array}
\right\vert
\begin{array}
[c]{cccc}%
0 & 0 & 0 & 0\\
1 & 0 & 0 & 0\\
0 & 1 & 1 & 1\\
0 & 1 & 1 & 1
\end{array}
\right]  .
\end{equation}
Perform a CNOT\ from qubit two to qubit three and from qubit two to qubit
four. The matrix becomes%
\begin{equation}
\left[  \left.
\begin{array}
[c]{cccc}%
1 & 0 & 0 & 0\\
0 & 0 & 0 & 0\\
0 & 1 & 1 & 0\\
0 & 1 & 0 & 1
\end{array}
\right\vert
\begin{array}
[c]{cccc}%
0 & 0 & 0 & 0\\
1 & 0 & 0 & 0\\
0 & 1 & 0 & 0\\
0 & 1 & 0 & 0
\end{array}
\right]  .
\end{equation}
Perform a phase gate on qubit two:%
\begin{equation}
\left[  \left.
\begin{array}
[c]{cccc}%
1 & 0 & 0 & 0\\
0 & 0 & 0 & 0\\
0 & 0 & 1 & 0\\
0 & 0 & 0 & 1
\end{array}
\right\vert
\begin{array}
[c]{cccc}%
0 & 0 & 0 & 0\\
1 & 0 & 0 & 0\\
0 & 1 & 0 & 0\\
0 & 1 & 0 & 0
\end{array}
\right]  .
\end{equation}
Perform a Hadamard on qubit three followed by a CNOT\ from qubit two to qubit
three:%
\begin{equation}
\left[  \left.
\begin{array}
[c]{cccc}%
1 & 0 & 0 & 0\\
0 & 0 & 0 & 0\\
0 & 0 & 0 & 0\\
0 & 0 & 0 & 1
\end{array}
\right\vert
\begin{array}
[c]{cccc}%
0 & 0 & 0 & 0\\
1 & 0 & 0 & 0\\
0 & 1 & 0 & 0\\
0 & 1 & 1 & 0
\end{array}
\right]  .
\end{equation}
Add row three to row four and perform a Hadamard on qubit two:%
\begin{equation}
\left[  \left.
\begin{array}
[c]{cccc}%
1 & 0 & 0 & 0\\
0 & 0 & 0 & 0\\
0 & 1 & 0 & 0\\
0 & 0 & 0 & 1
\end{array}
\right\vert
\begin{array}
[c]{cccc}%
0 & 0 & 0 & 0\\
1 & 0 & 0 & 0\\
0 & 0 & 0 & 0\\
0 & 0 & 1 & 0
\end{array}
\right]  .
\end{equation}
Perform a Hadamard on qubit four followed by a CNOT\ from qubit three to qubit
four. End by performing a Hadamard on qubit three:%
\begin{equation}
\left[  \left.
\begin{array}
[c]{cccc}%
1 & 0 & 0 & 0\\
0 & 0 & 0 & 0\\
0 & 1 & 0 & 0\\
0 & 0 & 1 & 0
\end{array}
\right\vert
\begin{array}
[c]{cccc}%
0 & 0 & 0 & 0\\
1 & 0 & 0 & 0\\
0 & 0 & 0 & 0\\
0 & 0 & 0 & 0
\end{array}
\right]  .
\end{equation}
The above matrix now corresponds to the canonical Paulis
(\ref{eq:canonical-Paulis}). Adding one half of an ebit to the receiver's side
gives the canonical stabilizer (\ref{eq:canonical-stabilizer}) whose
simultaneous +1-eigenstate is (\ref{eq:canonical-state}).

Figure~\ref{fig:encoding-circuit-science} gives the encoding circuit
corresponding to the above operations. The above operations in reverse order
take the canonical stabilizer (\ref{eq:canonical-stabilizer}) to the encoded
stabilizer (\ref{eq:encoded-stabilizer}).

\bibliographystyle{IEEEtran}
\bibliography{cedc-ieee}

\begin{thebibliography}{10}
\providecommand{\url}[1]{#1}
\csname url@samestyle\endcsname
\providecommand{\newblock}{\relax}
\providecommand{\bibinfo}[2]{#2}
\providecommand{\BIBentrySTDinterwordspacing}{\spaceskip=0pt\relax}
\providecommand{\BIBentryALTinterwordstretchfactor}{4}
\providecommand{\BIBentryALTinterwordspacing}{\spaceskip=\fontdimen2\font plus
\BIBentryALTinterwordstretchfactor\fontdimen3\font minus
  \fontdimen4\font\relax}
\providecommand{\BIBforeignlanguage}[2]{{%
\expandafter\ifx\csname l@#1\endcsname\relax
\typeout{** WARNING: IEEEtran.bst: No hyphenation pattern has been}%
\typeout{** loaded for the language `#1'. Using the pattern for}%
\typeout{** the default language instead.}%
\else
\language=\csname l@#1\endcsname
\fi
#2}}
\providecommand{\BIBdecl}{\relax}
\BIBdecl

\bibitem{PhysRevA.52.R2493}
P.~W. Shor, ``Scheme for reducing decoherence in quantum computer memory,''
  \emph{Phys. Rev. A}, vol.~52, no.~4, pp. R2493--R2496, Oct 1995.

\bibitem{PhysRevA.54.1098}
A.~R. Calderbank and P.~W. Shor, ``Good quantum error-correcting codes exist,''
  \emph{Phys. Rev. A}, vol.~54, no.~2, pp. 1098--1105, Aug 1996.

\bibitem{PhysRevLett.77.793}
A.~M. Steane, ``Error correcting codes in quantum theory,'' \emph{Phys. Rev.
  Lett.}, vol.~77, no.~5, pp. 793--797, Jul 1996.

\bibitem{thesis97gottesman}
D.~Gottesman, ``Stabilizer codes and quantum error correction,'' Ph.D.
  dissertation, California Institue of Technology, 1997.

\bibitem{ieee1998calderbank}
A.~Calderbank, E.~Rains, P.~Shor, and N.~Sloane, ``Quantum error correction via
  codes over gf(4),'' \emph{IEEE Trans. Inf. Theory}, vol.~44, pp. 1369--1387,
  1998.

\bibitem{book1983code}
F.~J. MacWilliams and N.~J.~A. Sloane, \emph{The Theory of Error-Correcting
  Codes}.\hskip 1em plus 0.5em minus 0.4em\relax North Holland, 1983.

\bibitem{science2006brun}
T.~A. Brun, I.~Devetak, and M.-H. Hsieh, ``Correcting quantum errors with
  entanglement,'' \emph{Science}, vol. 314, no. 5798, pp. pp. 436 -- 439,
  October 2006.

\bibitem{arx2006brun}
------, ``Catalytic quantum error correction,'' \emph{arXiv:quant-ph/0608027},
  August 2006.

\bibitem{unpub2007got}
\BIBentryALTinterwordspacing
D.~Gottesman. [Online]. Available:
  \url{http://www.perimeterinstitute.ca/personal/dgottesman/QECC2007/Sols9.pdf}
\BIBentrySTDinterwordspacing

\bibitem{PhysRevLett.76.722}
C.~H. Bennett, G.~Brassard, S.~Popescu, B.~Schumacher, J.~A. Smolin, and W.~K.
  Wootters, ``Purification of noisy entanglement and faithful teleportation via
  noisy channels,'' \emph{Phys. Rev. Lett.}, vol.~76, no.~5, pp. 722--725, Jan
  1996.

\bibitem{PhysRevA.54.3824}
C.~H. Bennett, D.~P. DiVincenzo, J.~A. Smolin, and W.~K. Wootters,
  ``Mixed-state entanglement and quantum error correction,'' \emph{Phys. Rev.
  A}, vol.~54, no.~5, pp. 3824--3851, Nov 1996.

\bibitem{PhysRevLett.69.2881}
C.~H. Bennett and S.~J. Wiesner, ``Communication via one- and two-particle
  operators on einstein-podolsky-rosen states,'' \emph{Phys. Rev. Lett.},
  vol.~69, no.~20, pp. 2881--2884, Nov 1992.

\bibitem{PhysRevLett.70.1895}
C.~H. Bennett, G.~Brassard, C.~Cr\'epeau, R.~Jozsa, A.~Peres, and W.~K.
  Wootters, ``Teleporting an unknown quantum state via dual classical and
  einstein-podolsky-rosen channels,'' \emph{Phys. Rev. Lett.}, vol.~70, no.~13,
  pp. 1895--1899, Mar 1993.

\bibitem{PhysRevLett.85.441}
P.~W. Shor and J.~Preskill, ``Simple proof of security of the bb84 quantum key
  distribution protocol,'' \emph{Phys. Rev. Lett.}, vol.~85, no.~2, pp.
  441--444, Jul 2000.

\bibitem{book2000mikeandike}
M.~A. Nielsen and I.~L. Chuang, \emph{Quantum Computation and Quantum
  Information}.\hskip 1em plus 0.5em minus 0.4em\relax Cambridge University
  Press, 2000.

\bibitem{luo:010303}
Z.~Luo and I.~Devetak, ``Efficiently implementable codes for quantum key
  expansion,'' \emph{Phys. Rev. A}, vol.~75, no.~1, p. 010303, 2007.

\bibitem{PhysRevLett.91.177902}
H.~Ollivier and J.-P. Tillich, ``Description of a quantum convolutional code,''
  \emph{Phys. Rev. Lett.}, vol.~91, no.~17, p. 177902, Oct 2003.

\bibitem{arxiv2004olliv}
------, ``Quantum convolutional codes: fundamentals,''
  \emph{arXiv:quant-ph/0401134}, 2004.

\bibitem{isit2005forney}
J.~G.~David~Forney and S.~Guha, ``Simple rate-1/3 convolutional and tail-biting
  quantum error-correcting codes,'' in \emph{IEEE International Symposium on
  Information Theory (arXiv:quant-ph/0501099)}, 2005.

\bibitem{ieee2007forney}
G.~D. Forney, M.~Grassl, and S.~Guha, ``Convolutional and tail-biting quantum
  error-correcting codes,'' \emph{IEEE Trans. Inf. Theory}, vol.~53, pp.
  865--880, 2007.

\bibitem{book1999conv}
R.~Johannesson and K.~S. Zigangirov, \emph{Fundamentals of Convolutional
  Coding}.\hskip 1em plus 0.5em minus 0.4em\relax Wiley-IEEE Press, 1999.

\bibitem{PhysRevA.58.905}
H.~F. Chau, ``Quantum convolutional error-correcting codes,'' \emph{Phys. Rev.
  A}, vol.~58, no.~2, pp. 905--909, Aug 1998.

\bibitem{PhysRevA.60.1966}
------, ``Good quantum-convolutional error-correction codes and their decoding
  algorithm exist,'' \emph{Phys. Rev. A}, vol.~60, no.~3, pp. 1966--1974, Sep
  1999.

\bibitem{ieee2006grassl}
M.~Grassl and M.~R\"{o}tteler, ``Quantum convolutional codes: Encoders and
  structural properties,'' in \emph{Forty-Fourth Annual Allerton Conference},
  2006.

\bibitem{isit2006grassl}
------, ``Noncatastrophic encoders and encoder inverses for quantum
  convolutional codes,'' in \emph{IEEE International Symposium on Information
  Theory (quant-ph/0602129)}, 2006.

\bibitem{book2001symp}
A.~C. da~Silva, \emph{Lectures on Symplectic Geometry}.\hskip 1em plus 0.5em
  minus 0.4em\relax Springer, 2001.

\bibitem{arx2005dev}
I.~Devetak, A.~W. Harrow, and A.~Winter, ``A resource framework for quantum
  shannon theory,'' \emph{arXiv:quant-ph/0512015}, 2005.

\bibitem{itit1967viterbi}
A.~J. Viterbi, ``Error bounds for convolutional codes and an asymptotically
  optimum decoding algorithm,'' \emph{IEEE Trans. Inf. Theory}, vol.~13, pp.
  260--269, 1967.

\end{thebibliography}

\end{document}